\documentclass[amsmath,amssymb,aps,twocolumn,notitlepage,showpacs,pre,floatfix]{revtex4-1}
\usepackage{graphicx}
\usepackage{bm}
\usepackage{hyperref}
\usepackage{color}

\DeclareMathOperator{\sech}{sech}

\begin{document}

\title{Attractor non-equilibrium stationary states in perturbed
long-range interacting systems}

\author{Michael Joyce$^1$, Jules Morand$^{1,2,3,4}$,  and Pascal Viot$^4$}
\affiliation{$^1$ Laboratoire de Physique  Nucl\'eaire et de Hautes \'Energies, UPMC IN2P3 CNRS  UMR 7585, Sorbonne Universit\'es, 4, place Jussieu, 75252 Paris Cedex 05, France}
\affiliation{$^2$  
National Institute for Theoretical Physics (NITheP),Stellenbosch 7600, South Africa} 
\affiliation{$^3$ Institute of Theoretical Physics, Department of Physics, Stellenbosch University, Stellenbosch 7600, South Africa }
\affiliation{$^4$ Laboratoire de Physique
  Th\'eorique de la Mati\`ere Condens\'ee, UPMC, CNRS  UMR 7600, Sorbonne Universit\'es, 4, place Jussieu, 75252 Paris Cedex 05, France}

\begin{abstract}
Isolated long-range interacting particle systems appear generically to relax to non-equilibrium 
states (``quasi-stationary states" or QSS) which are stationary in the thermodynamic limit. 
A fundamental open question concerns the ``robustness" of these states when the system is 
not isolated. In this paper we explore, using both analytical and numerical approaches to a 
paradigmatic one dimensional model, the effect of a simple class of  perturbations. 
We call them  ``internal local perturbations" in that the particle energies are perturbed at 
collisions in a way which depends only on the local properties.  Our central finding is that the
effect of the perturbations is to drive all the very different QSS we consider towards
a unique QSS. The latter is thus independent of the initial conditions of the system,
but determined instead by both the long-range forces and the details of the perturbations 
applied. Thus in the presence of such a perturbation the long-range system evolves
to a unique non-equilibrium stationary state, completely different to its state in absence 
of the perturbation, and  it remains in this state when the perturbation is removed.
We argue that this result may be generic for long-range interacting systems
subject to  perturbations which are dependent on the local properties 
(e.g. spatial density or velocity distribution) of the system itself.
\end{abstract}
\date{\today}
\maketitle
\section{Introduction}

Systems  of large numbers of interacting particles are subject in their 
physical analysis to a fundamental distinction based on whether they
are short-range or long-range, depending on the rapidity of the
decay with separation of the two body interaction potential.
The distinction in its canonical form arises from the presence or absence 
of the property of {\it additivity} of the macroscopic energy, which plays 
a fundamental role in equilibrium statistical mechanics. 
While most of familiar laboratory systems studied in physics are short-range --- notably any
system constituted of neutral atoms or molecules --- there are numerous examples
also of long-range systems, ranging from self-gravitating systems
in astrophysics and cosmology, to vortices in turbulent fluids, laser cooled 
atoms, and even biological systems (for a review, see e.g.\cite{Campa2009,CDFR2014}).
Study of various isolated long-range  systems has shown that they evolve 
from generic initial conditions to microscopical non-Boltzmann equilibria, known
as ``quasi-stationary states" (QSS) because they evolve towards
the system's true statistical equilibrium on time-scales which 
diverge with the number of particles (see e.g. \cite{Yamaguchi2004,Joyce2010,PhysRevE.78.021130,Marcos2013a,PhysRevLett.105.210602,RochaFilho2014})
The evolution to such states appears not to be characteristic 
of all long-range interactions, but only of the sub-class of these interactions 
for which the pair force (rather than pair potential) is non-integrable at 
large distances (see \cite{PhysRevLett.105.210602,Gabrielli2014}).
As these systems remain in the QSS indefinitely
in the thermodynamic limit,  these states can be considered to be the
fundamental relevant macroscopic equilibria of such systems, 
just as Maxwell Boltzmann (MB) equilibria are for short-range
systems. Theoretically they are understood to be 
stationary solutions of the Vlasov equation which
in principle describes these systems in the relevant
thermodynamic limit. Unlike MB equilibria, they
are infinitely numerous at given values of the 
global conserved quantities, and the actual
equilibrium reached depends strongly on the
initial condition of the system.  

A basic question which arises about QSS in long-range interacting 
systems concerns the ``robustness" of such states. They
are strictly defined only for isolated Hamiltonian 
systems and the question is whether they continue
to exist when the system is not exactly isolated,
or exactly Hamiltonian, or both. For attempts to observe these intriguing 
equilibria in laboratory systems, which are necessarily perturbed by and 
coupled to the external world in some way, it is an essential to know 
whether these states can be expected to survive.
Studies of toy models coupled to a thermal bath
(see e.g. \cite{Chavanis2011}) show, unsurprisingly, that such 
a coupling sends
the system to its thermal equilibrium, on a time
scale which depends on the coupling.  However,
much more generally, it has been suggested 
on the basis of study of the one dimensional HMF model
(see e.g.\cite{Gupta2010PRL,Gupta2010b})
that QSS will disappear in the presence of any generic
stochastic perturbation to the dynamics. A study of
the effect of external stochastic fields with spatial correlation 
applied to the same model (see e.g.  \cite{Nardini2012,Nardini2012a}.)
shows however that interesting non-equilibrium steady states 
can be obtained in this case. 

In this article we explore the question of the 
robustness of QSS in long-range systems 
to weak perturbations using a paradigmatic toy model of
long-range interactions ---  a one dimensional
self-gravitating system  --- subjected to a
particular class of perturbations, which
we refer to as ``local internal perturbations":
the perturbations to the purely self-gravitating
dynamics occur when particles collide,
and are described by simple collision rules
for the colliding particles, which may be 
stochastic or deterministic. The perturbations
can thus be considered to model physical effects 
which come into play at very small scales e.g. due 
to very short scale forces and/or internal 
degrees of freedom. Differently to the
study of  \cite{Nardini2012,Nardini2012a} there are
therefore no external forces, and indeed
we will build our collision rules so 
that they conserve total momentum. 
The choice of the two specific two models 
we study is then guided by simplicity:
once momentum  conservation is imposed, a 
non-trivial collision rule in one dimension
cannot conserve energy (as an elastic
collision gives rise simply to exchange
of particle velocities).  As we wish to
focus here on the effect on 
the effect of the perturbations on 
QSS, which have
fixed energy, we choose to define
perturbations which can conserve 
average energy and lead (in principle)
to a steady state. The two models
we study are then simple choices
with this property, corresponding to
collision rules drawn from the literature
on granular gases arising from simple
considerations of energy balance.
In the one dimensional self-gravitating model 
these perturbations given by non-trivial
collision rules are both very simple to 
implement numerically, and, as we
will see, also admit a straightforward 
theoretical description for the kinetic
theory in an appropriate  mean-field
limit. 

Our main result is that in both models we 
study we observe an evolution from the
initial QSS, which arises from a given initial
condition, through a family of states, until
a truly stationary state is reached. The
evolution is through a continuum of QSS
and the final state is also a QSS: if the
perturbation is removed the system remains
in this state. Further the final stationary state
in each case appears to be an attractor for
the perturbed long-range system, i.e., 
starting from different initial conditions which
evolve to very different QSS the perturbation
drives them all finally to the same non-equilibrium
state.  This ``universal'' QSS  is thus determined
essentially by the detailed nature of the small
perturbation and the long-range force itself.
 
The paper is organized as follows. We first define
the two models we study of self-gravitating particles 
perturbed in a specific manner when particles 
collide. In the following section
we describe analytical approaches to these
models which are valid in an appropriate 
mean-field limit. These give rise to kinetic
equations which allow us to determine a
well defined large $N$ limit. They also 
provide predictions for the early time
evolution of the system.  In the next section 
we present results of
numerical studies of the two models.
In our conclusions we compare our
results with other relevant works in
the literature, and comment on the 
possible generality of the behaviours 
we observe in a broad class of
perturbed long-range systems.  

\section{Models}

\subsection{The sheet model}

We consider  a system of  identical particles of mass $m$ moving in a one dimensional space
and interacting by a force independent of their separation, i.e. the force on a particle $i$ due 
to a particle $j$ is 
\begin{equation}
 F_{ij} =  −gm^2 sgn(x_i-x_j),
\end{equation}
where $g$ is the interaction strength. The model is known as the ``sheet model"
because these particles in one dimension are equivalent to infinite, infinitely thin, 
parallel sheets moving in three dimensions interacting by Newtonian gravity,  in 
which case $gm = 2\pi\Sigma G$, where $G$ is Newton's constant 
and $\Sigma$ is the mass per unit surface area of the sheets.  This
model dates back at least to the early study of  Camm\cite{Camm1950}
and has been studied quite extensively by numerous authors since (see e.g. \cite{Miller1996,Yawn1997,Joyce2011a}  and references therein). 
 
For a finite system of $N$ particles,  the total force acting on the $i$th particle at any time 
is simply given as \begin{equation}
 F_i = gm^2[N_i^+ - N_i^-],
\end{equation}
where $N_i^+$  and $N_i^-$ denote the numbers of particles on the right and on the left 
of $i$th particle, respectively.

The dynamics of this system has been extensively studied (see e.g. \cite{Miller1996,Yawn1997,Joyce2011a})
and shows that the time evolution of the system displays long-lived QSS before 
reaching equilibrium. More precisely the relaxation time associated with QSS is diverging 
with the system size $\tau\sim A N$, where the coefficient $A$ depends strongly on the initial states
 \cite{Joyce2011a}.

We consider now two variants of this model, in which the particle collision still conserves the momentum
but not the total energy. On the other hand, as we wish the system to be able to attain a
stationary state at constant energy, we constrain the exact collision rules to allow this.
Both collision laws are taken from simple models of granular systems which have been
studied in the literature (see references below).

\subsection{Model A: collisions with random coefficients of restitution}

Let us denote $v_{ij}=v_i-v_j$ the relative velocity of particles $i$ and $j$ which undergo
a collision with precollisional velocities $v_i$ and $v_j$. We adopt the rule that 
the postcollisional velocities, $v^*_i$ and $v^*_j$, are given by 
\begin{align}\label{eq:SCM}
v_{i}^{*}= & v_{j}+\frac{1-c}{2}  v_{ij}\\ \nonumber
v_{j}^{*}= & v_{i}- \frac{1-c}{2}  v_{ij} ,
\end{align}
where the coefficient of restitution $c$ is  a { \it non-negative} random variable.
Equivalently it corresponds to momentum conservation combined with the
rule 
\begin{align}
\label{eq:SCM-relative}
v_{ij}^{*} = - c v_{ij} \,.
\end{align}
Granular models of this kind, incorporating a random coefficient of restitution, have 
been introduced by ~\cite{Barrat2001} in order to include  the effect of energy injection 
in a vibrated two-dimensional  granular gas.

In each collision the change of the kinetic energy is given by 
\begin{equation}
\label{dK-A}
\delta K =\frac{m}{4} [( v_{ij}^*)^{2} - ( v_{ij})^{2}]  = \frac{c^2-1}{4} ( v_{ij})^{2} 
\end{equation} 
i.e. the collision is inelastic if $0 \leq c < 1$ and super-elastic if $c>1$.
In order that the applied perturbation may admit stationary states, we 
choose $c$ from a bimodal probability distribution in which $c$ takes two 
values, $c_A$ and $\tilde{c}_A$, with equal weight, with $0<c_A<1$ and 
\begin{equation}
\label{energy-con-condition-modA}
\tilde{c}_A=\sqrt{2-c_A^{2}}
\end{equation}

The latter relation imposes that, for a collision
at the same initial relative velocity, the energy lost with $c=c_A$
is the same as the energy gained when $c=\tilde{c}_A$. 
Thus, in the ensemble of realizations of the stochastic 
perturbations, the average energy is constant, with the 
energy loss of the inelastic collisions ($0<c_A<1$) balanced by the
energy gain in superelastic collisions  ($\tilde{c}_A > 1$).
We expect that in this case the system may be able
to reach a stationary state with constant energy
(modulo finite $N$ fluctuations).

\subsection{Model B:  inelastic collisions with energy injection}

In this model the self-gravitating particles undergo collisions specified by the 
following rule:
\begin{align}\label{eq:BSRcollision}
v_{i}^{*}= & v_{j}+\frac{1-c_B}{2} v_{ij} - \epsilon_{ij}\Delta \\ \nonumber
v_{j}^{*}= & v_{i}-\frac{1-c_B}{2} v_{ij} +\epsilon_{ij}\Delta
\end{align}
where the constant $c_B$ has a fixed value in the range $0 < c_B <1$
(i.e. as for an inelastic collision), $\Delta$ is a positive constant
(with dimensions of velocity) and $\epsilon_{ij}=sgn (v_{ij})$.  
The collision manifestly still conserves total momentum, 
and corresponds to 
\begin{align}
\label{eq:B-relative}
v_{ij}^{*} = - sgn (v_{ij})  [c_B |v_{ij}| + 2\Delta] \,.
\end{align}
and therefore the energy change is
\begin{eqnarray}
\label{eq:BRSdeltaK}
\delta K
=m\left[\frac{c_B^2-1}{4}v^2_{ij} + c_B \Delta | v_{ij} |+\Delta^2\right] 
\end{eqnarray}

The term in $\Delta$ in the collision rule thus leads to 
an energy injection, which can be smaller or larger than the
energy loss due to the inelastic term:  more
precisely, the collision leads to an energy
gain if $|v_{ij}| < v_0$, and an energy loss
if $|v_{ij}| > v_0$, where  
\begin{equation}
\label{eq:v0}
v_0=\frac{2\Delta}{1-c_B}
\end{equation}
is the value of the relative velocity for which 
the collision is elastic.

This collision rule is the one dimensional version of that introduced in two dimensions by \cite{Brito2013} 
in a phenomenological model of quasi-two dimensional experiments  of agitated granular particles:
the particles  are confined  between two horizontal plates, and the vibrating bottom plate  transfers the kinetic 
energy to the particles by collisions~\cite{Olafsen2005,Rivas2011,Neel2014}.  In this quasi-two dimensional geometry, the period  
of the vertical vibrations is much shorter than the typical time scale of the horizontal dynamics. The collision 
rule, Eq.~(\ref{eq:BSRcollision}), then represents a time coarse-grained description of the energy transfer of 
particle-bottom plate collisions to  horizontal particle-particle collisions. 

In this paper we have chosen this collision rule simply because it provides a simple way, quite different to 
that in the first model, to obtain a non-trivial two body collision rule which can be expected to lead to 
a stationary state. More specifically if the particle velocities at collisions are assumed to be 
uncorrelated, the kinetic energy of the system gives a direct measure of the typical relative 
velocity of colliding particles: $\langle (v_i-v_j)^2\rangle=2 \langle v_i^2\rangle$. The kinetic energy 
would then be expected to be driven towards a value of order $Nv_0^2$, as above this 
energy scale energy will be dissipated while below it energy will be injected. Indeed
in the case in which gravity is turned off, and the particles are enclosed in a box with
reflecting walls, if all particles have velocity $\pm v_0/2$ all collisions are elastic
and the velocity distribution does not evolve at all.  

Model B is in fact  microscopically deterministic, while Model A is explicitly stochastic.
One other notable difference is that  the phase space volume occupied by
particles involved in a collision strictly contracts in Model B, while it
can contract or increase in Model A depending on whether the 
collision is inelastic or elastic. Indeed for a collision with
coefficient of restitution $c$ in either model we have
$dv^*_idv^*_j=cdv_idv_j$ which is always
a contraction in Model B. This property leads to distinctive features 
of the long time behaviour of the models which we observe below.

\section{Kinetic Theory}

\subsection{Mean field limit without collisions}
For the purely self-gravitating model the dynamics in the appropriate large $N$ 
mean field limit is described by the Vlasov equation \cite{Chavanis2010,Bouchet2010, Gabrielli2014}.
This limit is obtained  by taking $N \rightarrow \infty$ at fixed values
of the total system mass $M$ and energy $E$. 
Denoting the mass density in phase space,  $f(x,v,t)$,  the Vlasov equation reads
\begin{align}\label{eq:bve1}
 \partial_t f(x,v,t)+J_V[f]=0
 \end{align}
 where $J_V[f]$ the Vlasov operator is
 \begin{align}
J_V[f]= v\partial_x f(x,v,t) +
\bar{a}(x,t) \partial_v f(x,v,t)
\end{align}
where $\bar{a}(x)$ is the mean-field acceleration given by
$
\bar{a}(x,t)  =  g \int sgn(x-x') f(x',v',t)dx'dv'
$.
The mass density obeys the normalization condition
\begin{equation}
 \int \int dx dv f(x,v,t)=M\,.
\end{equation}
where $M$ is the total mass of the system.

QSS are interpreted as stationary solutions of Eq.(\ref{eq:bve1}). There
are an infinite number of such solutions, including 
as a particular case the statistical equilibrium of this model 
(see below).

\subsection{Model A}\label{sec:modelI}

\subsubsection{Collision operator in Boltzmann approximation}
The Vlasov equation can be derived starting from the BBGKY
hierarchy and making the approximation that the two point 
correlations can be neglected. The collisions in our model
can be treated in the same approximation, and are then
described by a canonical Boltzmann operator.  
We thus expect our model in the mean field limit
to be described by a kinetic equation
\begin{align}\label{eq:Vlasov_SCM}
&\partial_t f(x,v,t)+v.\partial_x f(x,v,t) \nonumber\\
&+\partial_x \bar{a}[f](x,t) .\partial_v f(x,v,t)= \sum_q P(q) J_{q}[f,f](x,v,t) \, . 
\end{align}
where, for convenience, we introduce the parameter $q=\frac{1-c}{2}$
to characterize binary collisions with a coefficient of restitution equal to $c$, 
and $J_{q}[f,f](x,v,t)$ is a collision operator accounting for such
collisions which are assumed to occur independently with a probability 
$P(q)$. Initially we will leave $P(q)$ undetermined and then replace 
it with the specific bimodal form for Model A at the appropriate
point below.

Assuming the particles to be pointlike, the collision operator is a homogeneous 
Boltzmann operator accounting for binary collisions, which is the sum of 
two contributions:
\begin{equation}\label{eq:balance}
J_{q}[f,f](x,v,t)=G_{q}(x,v,t)-L(x,v,t)\, .
\end{equation} 
where $G_q$ is the gain term corresponding to collisions where a particle has
a post-collisional velocity equal to $v$,   
\begin{align}
\label{gain-term-1}
G_{q}(x,v,t) &=\frac{N}{M} \int \int dv'dv''|v'-v''|f(x,v',t)f(x,v'',t)\nonumber\\
&\delta(v-q v'-(1-q) v'') \,, 
\end{align}
and $L(x,v,t)$ is the loss term corresponding to collisions where a particle with a 
velocity $v$ undergoes a collision at time $t$, 
\begin{equation}
L(x,v,t)=\frac{N}{M}f(x,v,t)\int dv' |v'-v|f(x,v',t) \, .
\end{equation}
Note that the loss term does not depend explicitly on the coefficient of restitution,
and indeed we can write  
\begin{equation}
J_{q}[f,f](x,v,t)=G_{q}(x,v,t) - G_{0}(x,v,t)
\end{equation}

Let us introduce a series expansion of the $\delta$ function in terms of the parameter $q$ 
\begin{equation}
 \delta(v-q v'-(1-q) v'')=\sum_{n\geq 0}\frac{(q(v''-v'))^n}{n!}\delta^{(n)}(v-v'')
\end{equation}
where $\delta^{(n)}$ denotes the $n$th derivative of the $\delta$ function.

The Boltzmann operator is then expressed as
\begin{align}\label{eq:BoltzmannSTOCH}
J_{q}[f,f](x,v,t)=&\frac{N}{M} \sum_{n\geq 1} \!\int\!\!\! \int\!\!
dv'dv''|v'-v''|\frac{q^n((v''-v'))^n}{n!}\nonumber \\
&\delta^{(n)}(v-v'') f(x,v',t)f(x,v'',t)
\end{align}

To determine whether the parameters characterizing the collisions
can be rescaled with $N$ so that the collision term remains well
defined (and non-trivial) in the mean-field (Vlasov) limit, we consider 
the limit $q \rightarrow 0$ of the model i.e. the {\it quasi-elastic} limit. 
Physically this is clearly the relevant limit: to
obtain an $N$ independent evolution in presence of the
collisions on time-scales characterizing the mean-field dynamics 
(e.g. the time a particle typically takes to cross the system)
one must clearly ``compensate'' the effect of the divergent
growth of the number  of collisions with $N$ by making 
the effect of each collision arbitrarily weak.

\subsubsection{Expansion of kinetic equation}
Inserting Eq.~(\ref{eq:BoltzmannSTOCH}) in the right hand side of  Eq.~(\ref{eq:Vlasov_SCM})
and integrating by parts term by term we obtain (following \cite{McNamara1993,Talbot2011,Joyce2014})
the collision operator given as a series of differential operators:
\begin{align}\label{eq:fokkerplanck}
J_{A}[f]&=\frac{N}{ M} \sum_{n\geq 1} \left[\langle q^n \rangle\partial_v^n(f(x,v,t)a_n[f](x,v,t))\right.\nonumber\\
\end{align}
with  
\begin{equation}\label{eq:an}
a_n[f](x,v,t)=\int|v'-v|\frac{(v'-v)^n}{n!}f(x,v',t)dv'
\end{equation}
where the brackets $\langle..\rangle$ indicate an average over 
the probability distribution $P(q)$. Note that $a_1[f](x,v,t)$ has a simple physical 
meaning as an average effective force due to the collisions 
\cite{PhysRevE.82.011135,Fruleux2012}.

Let us consider now the specific $P(q)$ of model A:
\begin{equation}
P(q)=\frac{1}{2} \delta(q-q_A) + \frac{1}{2} \delta(q-\tilde{q}_A)
\end{equation}
where $q_A=\frac{1-c_A}{2}$ and, from
the average energy conserving condition Eq.(\ref{energy-con-condition-modA}), 
\begin{equation}
\tilde{q}_A=\frac{1-\sqrt{1+4(1-q_A)q_A}}{2}
\end{equation}
Expanding in ${q}_A$ (as $q_A \rightarrow 0$ in the 
quasi-elastic limit) we have
\begin{equation}
\tilde{q}_A=-{q}_A + 2 {q}_A^2 - 4{q}_A^3 + O({q}_A^4)
\end{equation}
and thus, to leading order in powers of ${q}_A$ we 
obtain
\begin{align}
\langle q^n\rangle = &n{q}_A^{n+1} \quad n \, {\rm odd} \nonumber \\
\langle q^n\rangle = &{q}_A^{n} \quad \quad  n \, {\rm even} \,. \nonumber
\end{align} 
Defining now
\begin{equation}
\label{gammaA_def}
{\gamma}_A \equiv  q_A \, \sqrt{N}= \frac{(1-c_A)\sqrt{N}}{2}
\end{equation}
we have, at leading order in $1/N$,  
\begin{align}
\langle q\rangle =\langle q^2\rangle = \frac{\gamma_A^2}{N}\nonumber\\
\langle q^3\rangle= 3 \langle q^4 \rangle =\frac{3\gamma_A^4}{N^2} \,, \nonumber
\end{align}
while, for $n>4$,  $\langle q^n\rangle$ decreases with $N$ more rapidly than $1/N^2$. 
Thus taking the mean-field limit $N \rightarrow \infty$
at constant $\gamma_A$, the full expansion of the collision term
reduces to the sum of the two first derivatives of $f(x,v,t)$:
\begin{align}
\label{collision-term-mean-field}
J_{A}[f]=\frac{\gamma_A^2}{ M}  &\left[\partial_v(f(x,v,t)a_1[f](x,v,t))\right.\nonumber\\
&\left.+  \partial_v^2(f(x,v,t)a_2[f](x,v,t))   \right]
\end{align}
Note that the non-linear structure of the integral collision operator Eq.~(\ref{eq:fokkerplanck}) is 
conserved because the velocity-dependent functions $a_n(v)$ are functionals of 
$f(x,v,t)$.  

We note that the crucial relation leading to the result
 (\ref{collision-term-mean-field}) for this model 
is $\langle q \rangle=\langle q^2\rangle$, which is 
simply the condition of average energy conservation. 
Indeed from Eq. (\ref{dK-A}) it follows that the energy change
in a collision at any given relative velocity is proportional 
to $q- q^2$. Thus the same mean-field limit for the
kinetic theory will be obtained for any variant 
of this model in which $P(q)$ is such that
energy is conserved on average.

We note further that in this derivation we have assumed implicitly that 
$f(v)$ has the convergence properties required for the validity of the Taylor expansion, 
which requires clearly sufficiently rapid decay of $f(v)$ at large $v$ to ensure 
the finiteness of the coefficients. Indeed we see that while the
finiteness of Eq.(\ref{gain-term-1}) requires only that $f(v)$
be integrable at large $|v|$ (i.e $f(v) \sim 1/|v|^\alpha$ with
$\alpha >2$), the definiteness of the expression 
Eq.(\ref{collision-term-mean-field}) requires $\alpha >4$.
As we will discuss below the latter assumption
turns out to break down at longer times in the 
model. 

\subsubsection{Evolution of moments of velocity distribution}
We now discuss some properties of the collision operator by considering the 
evolution of the moments of the velocity distribution. Multiplying both
sides of the kinetic equation by $v^n$, and integrating over $v$,
we obtain 
\begin{equation}
\label{eq:SCMop}
 \frac{d}{dt} [\rho (x,t) \overline{v^n}  (x,t)]  + \int dv \, v^n J_V[f]=\int dv \, v^nJ_{A}[f]
\end{equation}
where $\rho(x,t)=\int dv f(x,v,t)$ is the spatial mass density, and 
$\overline{v^n}  (x,t)$ is the $n$-th moment of the velocity
distribution at $x$, i.e.,  
\begin{equation}
\label{eq:vel-moments}
\overline{v^n}  (x,t) = \int dv \, v^n p_x (v,t) \quad {\rm where} \quad p_x (v,t)= \frac{f(x,v,t)}{\rho(x,t)}
\end{equation}

It is straightforward to show, either directly from the exact collision 
operator,  or for each of the two terms in Eq.~(\ref{eq:SCMop}), 
that 
\begin{eqnarray}
\label{first-moments}
\int dv J_{A}[f]=0 \nonumber \\
\int dv \, v J_{A}[f]=0 \nonumber
\end{eqnarray}
which express, respectively, the conservation of 
particle number and conservation of momentum in the collisions.
Indeed it is straightforward to show that  the left-hand side 
of Eq.~(\ref{eq:SCMop}) corresponds for $n=0$ to the 
continuity equation, and $n=1$ to the Euler equation. 

For the case $n=2$, integration by parts using
Eq.~(\ref{collision-term-mean-field}) gives
\begin{equation}
\label{eq:energy}
\int dv \,v^2 J_{A}[f]=2\frac{\gamma_A^2}{M}\int dv (v a_1[f]+a_2[f])f(x,v,t)
\end{equation}
from which it follows using Eqs. (\ref{eq:an})
that
\begin{equation}\label{eq:energy2}
\int dv \,v^2 J_{A}[f] =0 \,,
\end{equation}
which expresses the conservation of the kinetic energy
by the collisions. Thus the local pressure $\rho \overline{v^2}$
can change only due to the mean field gravitational
force (through the Vlasov flow term $J_V [f]$).
Note that while the vanishing of the 
zero and first moments hold for any $P(q)$,
it can be verified from Eq. (\ref{eq:fokkerplanck})
that the second moment vanishes only if
$\langle (q - q^2) \rangle = 0$ which is,
as noted above,  just  the condition of average energy 
conservation. 

For any $n \geq 2$ it is straightforward to show
that 
\begin{align}
\int &dv \, v^n J_{A}[f] =\frac{n\gamma_A^2}{4M}\int dv \int dv' |v-v'|(v-v')
  \nonumber\\
  & \times \left[(n-3)(v^{n-1}-v'^{n-1})-(n-1) v v' (v^{n-3}-v'^{n-3}) \right] \nonumber\\
 & \times f(x,v,t)f(x,v',t)
\end{align}
from which we recover the previous result for $n=2$, and further 
find that the first non-zero moment is for $n=4$, and it has the simple
expression
\begin{equation}
\label{eq:4thmoment}
\int dv \,v^4 J_{A}[f] = \frac{\gamma_A^2}{M} \int dv \int dv' |v-v'|^5f(x,v,t)f(x,v',t)\,.
\end{equation}

The fact that right hand of this expression is strictly positive has an important 
consequence: 
if this kinetic equation is valid, the system cannot reach a stationary state. 
Or, conversely, if the system reaches a stationary state, it must be such 
that the assumptions necessary for the derivation of the kinetic equation
break down. As noted above we will see that our numerical study shows 
that the system generically evolves to such a regime. In fact, we will see 
that when the system reaches a stationary state it is characterized by a 
non-Gaussian velocity distribution with tails decaying as a power law. 
Indeed the estimated exponent of the velocity distribution is such 
that the $4$th moment is not defined and thus the above equation 
is not applicable in this state.

On the other hand if the system is prepared in an initial state which
is a QSS, and which does satisfy the conditions
necessary for the validity of the derivation leading to
Eq. (\ref{collision-term-mean-field}), we can use 
Eq. (\ref{eq:4thmoment}) to infer non-trivial information
about the temporal evolution at sufficiently short times.
Indeed in this case
\begin{align}
\frac{d}{dt}\!\! \left(\!\rho (x,t) \overline{v^4} (x,t)\!\right)  \!\!=\!\! \frac{\gamma_A^2}{M}
\!\!\int\!\!\!dv\!\!\int\!\!\!dv' |v-v'|^5f(x,v,t)f(x,v',t)
\label{eq:4thmomentB}
\end{align}
In practice we measure the integrated quantity, i.e. the rescaled fourth moment of velocity (the kurtosis),
defined by  
\begin{align}\label{eq:kurtosis}
\beta_2 (t) = \frac{M \int dx \int dv \,v^4 f(x,v,t)}{\left(\int dx \int dv \,v^2 f(x,v,t)\right)^2} 
\end{align}
i.e. the fourth rescaled moment of the full velocity distribution $P(v,t)= \frac{1}{M}\int dx f(x,v)$.  
For a Gaussian distribution, the value of $\beta_2$ is constant, independant of the temperature of the system  and equal to $3$.
The time evolution $\beta_2(t)$ characterizes  deviation of the distribution from  a Gaussian distribution.

Integrating Eq. (\ref{eq:4thmomentB}) over $x$ and, assuming that the time evolution of the second moment (i.e. of the total kinetic energy) can be neglected, we have 
\begin{equation}
\label{kurtosis-evolutionKT}
\frac{d\beta_2(t)}{dt}\simeq  \frac{\gamma_A^2 \int dx \int dv \int dv'|v-v'|^5 f(x,v,t)f(x,v',t) }{\left(\int dx \int dv \,v^2 f(x,v,t)\right)^2 }
\end{equation}

We will test this prediction below for the case where the initial state is
the statistical equilibrium of the system.

\subsection{Model B}\label{sec:modelII}
Following exactly the same approach we write the kinetic equation,
in the Boltzmann and mean-field approximations,  for this model as 
\begin{align}\label{eq:Vlasov_NSR}
\partial_t f(x,v,t) + J_V[f]=
J_{B}[f,f](x,v,t) \, . 
\end{align}
where $J_{B}[f,f](x,v,t)$ is the collision operator 
with the same structure as Eq. ({\ref{eq:balance}), but with
the gain operator now given by 
\begin{align}
\label{gain-B-1}
G_{q}(x,v,t) &=\frac{N}{M} \int \int dv'dv''|v'-v''|f(x,v',t)f(x,v'',t)\nonumber\\
&\delta(v-q v'-(1-q) v''-sgn(v'-v'') \Delta) \,
\end{align}
where $q=\frac{1-c_B}{2}$ is a fixed positive parameter less than unity.

Following the same arguments as above, it is clear that to obtain a collision 
operator $J_{B}[f,f](x,v,t)$  which is independent of $N$ in the mean-field limit, we 
must consider a quasi-elastic limit,  with $q \rightarrow 0$ and $\Delta \rightarrow 0$
as $N \rightarrow \infty$. Further if the evolution to
a stationary state is to be described in such a limit, the energy of this
state, which we have inferred must be $\sim Mv_0^2$,
must be extensive (like the energy in the mean field
limit) and therefore $v_0$ must be taken independent of 
$N$. Now, since  $\Delta=qv_0$, holding $v_0$ fixed
and taking $q \rightarrow 0$ indeed defines a 
quasi-elastic limit.

Proceeding as in the previous case, we  perform again an expansion of the Boltzmann 
operator in powers of $q$ about $q=0$. This gives
\begin{equation}
 J_{B}[f,f]=\frac{N}{M} \sum_{n\geq 1}q^n\partial_{v}^n(a_n(x,v,t)f(x,v,t))
\end{equation}
where 
\begin{equation}\label{eq:a1BRS}
a_n[f](x,v,t)\!\!=\!\!\frac{1}{n!}\!\!\int\!\!|v'-v|(v'-v-sgn(v'-v)v_0)^n f(x,v',t)dv'
\end{equation}

Defining now 
\begin{equation}
{\gamma}_B\equiv q {N}=\frac{(1-c_B)N}{2}
\end{equation}
and taking $N\rightarrow \infty$ at constant $\gamma_B$,
we obtain a finite limit for the collision operator which
corresponds to the mean field limit. In this case only the leading 
linear term of the expansion contributes, and the effect of collisions 
corresponds to the presence of an effective
velocity dependent force per unit mass $a_1 (x,v)$:
\begin{align}
\label{eq:fokkerplanck2}
J_{B}[f]&=\frac{\gamma_B}{ M}  \left[\partial_v(f(x,v,t)a_1[f](x,v,t))\right]\,.
\end{align}
We note that the diffusive term which was non-zero in the 
mean-field limit of Model A thus vanishes for Model B.

As for the model A, we can calculate the velocity
moments of the collision operator  $J_{B}[f]$.
The first two moments again vanish 
as a consequence of conservation of particle number 
and momentum, while  
\begin{equation}\label{eq:energyB}
\int d vv^2  J_{B}[f]=2\frac{\gamma_B}{M}\int dv  v a_1[f]f(x,v,t)
\end{equation}
Differently to model A this is not zero, in general: indeed the
model does not necessarily conserve energy on average.
On the other hand we expect the system to be able to
reach a stationary state in which the collision operator is zero, 
and in this case Eq. (\ref{eq:energyB}) will vanish. 

\section{Numerical results}

\subsection{Simulation method and units}

\subsubsection{Code}

The molecular dynamics of a one dimensional self-gravitating model 
is conveniently simulated using an event-driven algorithm as 
between particle collisions trajectories can be calculated explicitly. 
Such an algorithm is exact up to the machine rounding error
in computing the solutions of quadratic equations giving
the collision times (see \cite{Miller1996,Joyce2011a,Joyce2011} and references therein). 
Further the algorithm may be sped up using a ``heap structure"  \cite{Noullez2003} and 
by updating positions at each step only the particles involved in each collision.
It is straightforward to modify this algorithm to implement the simple collision 
rules of our two models instead of elastic collisions (equivalent to particle
crossings). We use a modified version of the code described 
in \cite{Joyce2014} (and greater detail in \cite{Joyce2011})
\footnote{We note that this code uses periodic boundary conditions, 
which is equivalent to the presence of an additional repulsive 
force relative to the centre of mass, and of intensity proportional 
to the mean mass density in the box. This modification due
to the periodic boundary conditions is negligible when the region 
in which the particles move is very small compared to the box size.
This is true in all our simulations here, for which
the system size is typically one hundredth of the box size.}.
Model A is characterized by the choice of the parameter $c_A<1$,
and each collision is then chosen with probability $0.5$ to be inelastic
(with $c=c_A$) or superelastic (with $c=\tilde{c}_A$). For model B
is characterized fully by the values of $c_A$ and $\Delta$. 

\subsubsection{Initial conditions}

For both models we study evolution starting from three kinds of initial 
conditions:

\begin{itemize}
\item ``Rectangular waterbag":  particles are randomly distributed 
with uniform probability in a rectangular region of phase space, 
$[-L_0/2, L_0/2] \times$ $[-V_0/2, V_0/2]$. For the case of gravity only, 
which has no characteristic length scale, this is a one parameter family 
of initial conditions which may be conveniently characterized fully 
by the initial virial ratio $R_0$ (and the particle number $N$),
where the virial ratio $R$ (at any time) is defined as 
\begin{equation}\label{eq:defVR}
R=\frac{2K}{U}
\end{equation}
where $K$ is the kinetic energy and $U$ the potential
energy. A virialized system this has $R=1$.

\item  Thermal equilibrium: the statistical equilibrium of the purely
self-gravitating system in the micro-canonical and canonical
ensemble has been derived for any finite $N$ by \cite{Rybicki1971} and
its mean-field limit (derived earlier by \cite{Camm1950}) is
\begin{equation}
\label{eq:equivlasov}
 f(x,v)=\frac{M}{2\sqrt{\pi}\sigma\Lambda}e^{\frac{-v^2}{\sigma^2}}\sech^2\left(\frac{x}{\Lambda}\right)
\end{equation}
with
$\sigma^2=\frac{4E}{3M}$, $\Lambda=\frac{4E}{3gM^2}$
and $E$ is the total energy.

\end{itemize}
\subsubsection{Units} 

For our study the only dimensional parameters of relevance are the time,
and, in model B, the velocity (because of the parameter $\Delta$).  
A natural choice of units for both are those characteristic
of the mean field dynamics. For the time unit
we choose 
\begin{equation}
 \tau_{dyn}=\frac{1}{\sqrt{g\rho_0}}\,,
\label{def-dyntime}
\end{equation}
where $\rho_0$ is the initial mass density of the system, and
for the velocity
\begin{equation}
v_{dyn}= \sqrt{\frac{2E_0}{3M}} 
\label{def-dynvel}
\end{equation}
where $E_0$ is the initial energy. With this definition $v_{dyn}^2$
is the velocity dispersion of a virialized system with energy $E_0$.

Previous studies (see e.g. \cite{Joyce2011a}) of evolution from 
this first class of initial conditions for the self-gravitating system
show that the system evolves, in a time of order 
$10-100$ $\tau_{ dyn}$, to QSS of which the properties  
depend strongly on $R_0$. At longer times, 
of order $(10^2-10^3) N\tau_{dyn}$, the different QSS
all relax to thermal equilibrium
\cite{Joyce2010}.

\subsubsection{Additional macroscopic observables} 

To monitor in a simple way the evolution of the global
properties of the system, we measure in addition to
the energy and the virial ratio, that of a global parameter 
which is a simple measure of the ``phase space 
entanglement" of  the state of the system:
\begin{equation}
\phi_{11} = \frac{\langle |x v|\rangle }{\langle |x|\rangle\langle | v|\rangle}-1
\end{equation}
As shown in  \cite{Joyce2011a} the
only stationary solution of the Vlasov equation which
is a separable function of position and velocity  
is that corresponding to thermal equilibrium.
Thus if $\phi_{11}$ is constant and non-zero this indicates 
that the system  is in a QSS distinct from thermal equilibrium, 
and its amplitude can be taken roughly as  a measure 
of ``proximity" to the latter. We also monitor the evolution of the
kurtosis $\beta_2$ as defined in Eq. (\ref{eq:kurtosis}).

\subsection{Model A}

As seen in section~\ref{sec:modelI}, the relevant parameter characterizing 
the perturbations due to collisions in the mean-field limit is $\gamma_A$ defined in 
Eq. (\ref{gammaA_def}). If this limit describes accurately the dynamics
of the system, the term arising from the perturbations is 
proportional to $\gamma_A^2$, and thus the time scale 
on which they are expected to modify the evolution of the system 
is $\sim \tau_{dyn}/\gamma_A^2$. In order to study the desired range of 
weak perturbation, and assess the validity of the mean-field limit, we 
will thus consider small values of $\gamma_A$ and
vary $N$ keeping $\gamma_A$ fixed. 
We report here results for $\gamma_{A}=0.03$ and
$\gamma_{A}=0.1$, and for $N$ in the 
range $N=128$ to $N=1024$. We average our results 
over a large number of realizations in each case.

\subsubsection{Evolution of energy and virial ratio}

\begin{figure}[!h]
\begin{center}
\includegraphics[scale=0.37]{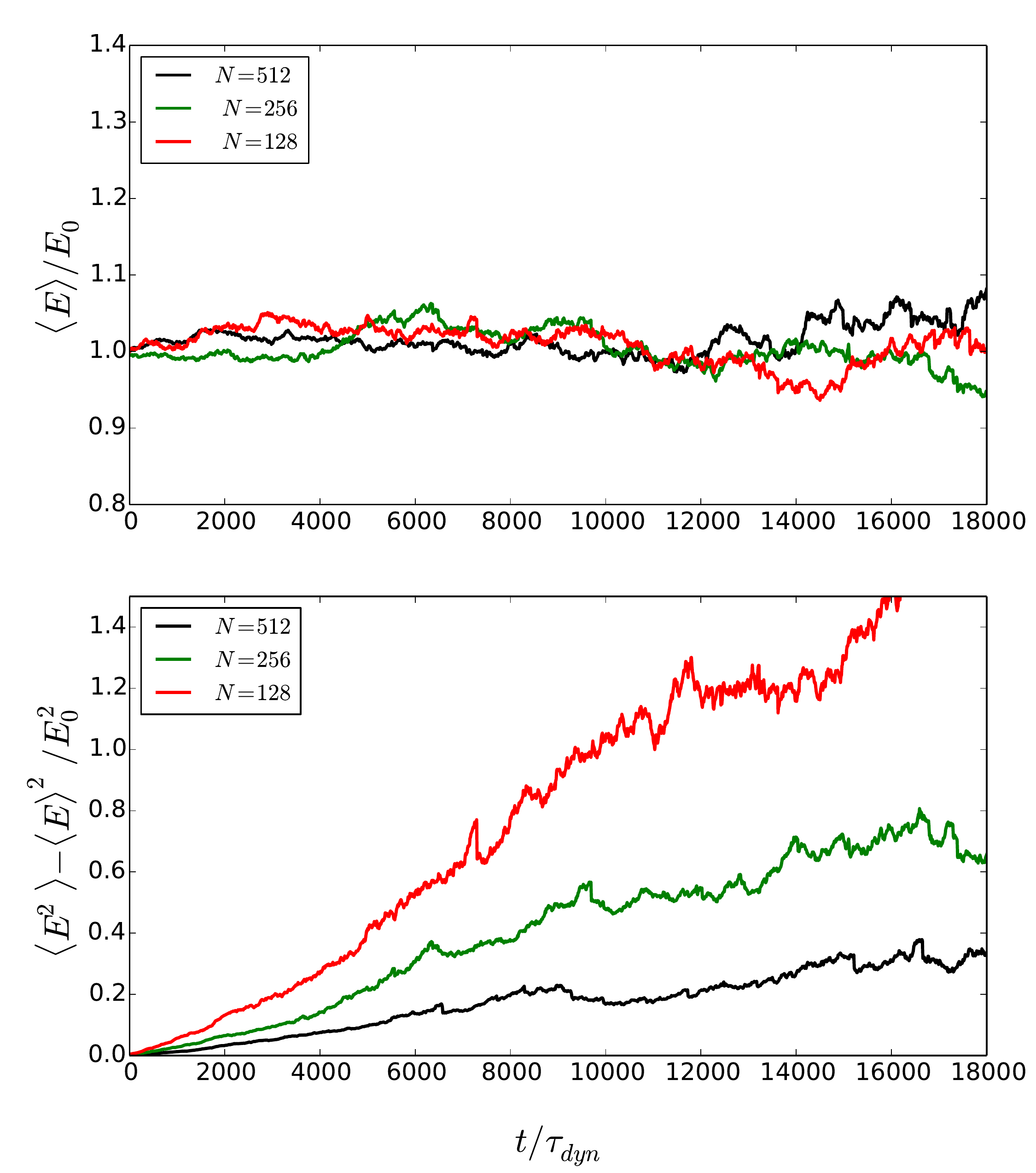}
\end{center}
\caption{Model A: (Top panel) Dimensionless energy $\langle E\rangle /E_0$  ($E_0$ is the initial energy) and (bottom panel) dimensionless variance 
$(\langle E^2\rangle -\langle E\rangle^2)/E_0^2$
versus dimensionless time $t/\tau_{dyn}$  averaged over $100$ realizations with $\gamma_{A}=0.03$, of rectangular waterbag initial conditions 
for  $N=128,256,512$ particles. } 
\label{fig:god_stoch_tot_all}
\end{figure}

We have constructed this model so that the fluctuations in energy
of the system should become arbitrarily small for sufficiently large $N$.
Indeed in the mean field limit we have derived above the energy
is exactly conserved, and this limit evidently thus does not 
describe effects associated with the energy fluctuations at finite $N$. 
In our numerical study, at finite $N$, we therefore need to check whether, 
on the time scale simulated, the energy
fluctuations are indeed small.  Fig.~\ref{fig:god_stoch_tot_all}  shows the evolution as a function 
of time of the mean energy, for a model with $\gamma_{A}=0.03$, in an ensemble 
of realizations starting from  waterbag initial conditions  with the different indicated $N$, over 
a time scale roughly an order of magnitude greater than $\tau_{dyn}/\gamma_A^2$. We see 
that, on these time scales, that in all cases, the ensemble averaged energy is 
indeed very close to constant, but (lower panel), the variance of the normalized energy
(i)  grows almost linearly in time, as indicated by the dashed straight lines
and (ii)  monotonically decreases as $N$ increases. Thus, as we would expect, 
taking $N$ sufficiently large at any given time, we can in principle converge to arbitrarily
precise conservation of the energy in any single realization. In
our numerical simulations at (relatively small) finite $N$, we have, 
however, significant finite $N$ fluctuations developing in all
cases at times a few times $\tau_{dyn}/\gamma_A^2$, so
we might anticipate that such effects may begin to play
a significant role on these time scales.

\begin{figure}[!h]
\begin{center}
\includegraphics[scale=0.37]{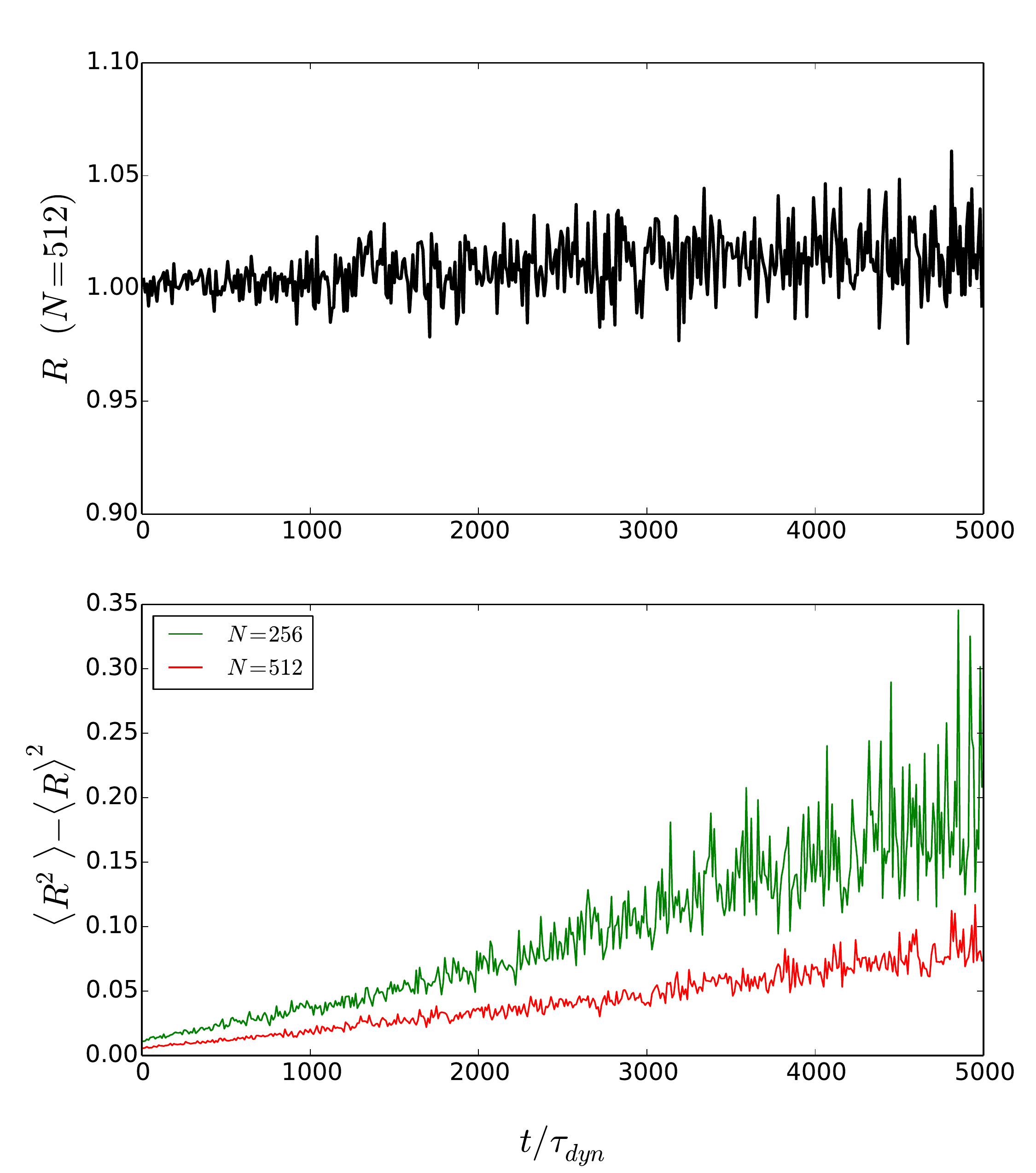}
\end{center}
\caption{Model A:  (top panel)  virial ratio $R$ as a function  of time $t/\tau_{dyn}$  averaged over
$100$ realizations with equilibrium initial conditions and for $\gamma_{A}=0.03$ and $N=512$. (Bottom panel) variance 
of the virial  ratio  for $N=512$ and $N=256$ with the
same initial conditions.}
\label{fig:god_stoch_vir_all}
\end{figure}

Fig.~\ref{fig:god_stoch_vir_all} shows the evolution of the mean (upper panel)
and standard deviation (lower panel) of the virial ratio in the same ensemble 
of simulations as in the previous figure. Because of the equilibrium initial  conditions, 
the system remains always, as we would expect, very close to virialized, with
only finite small $N$ fluctuations which (lower panel) clearly  
decrease monotonically as $N$ increases. We will see below that
the non-trivial time dependence of these fluctuations are a reflection
of the macroscopic evolution of the system on the same time scales.

\begin{figure}[h!]
\begin{center}
\includegraphics[scale=0.37]{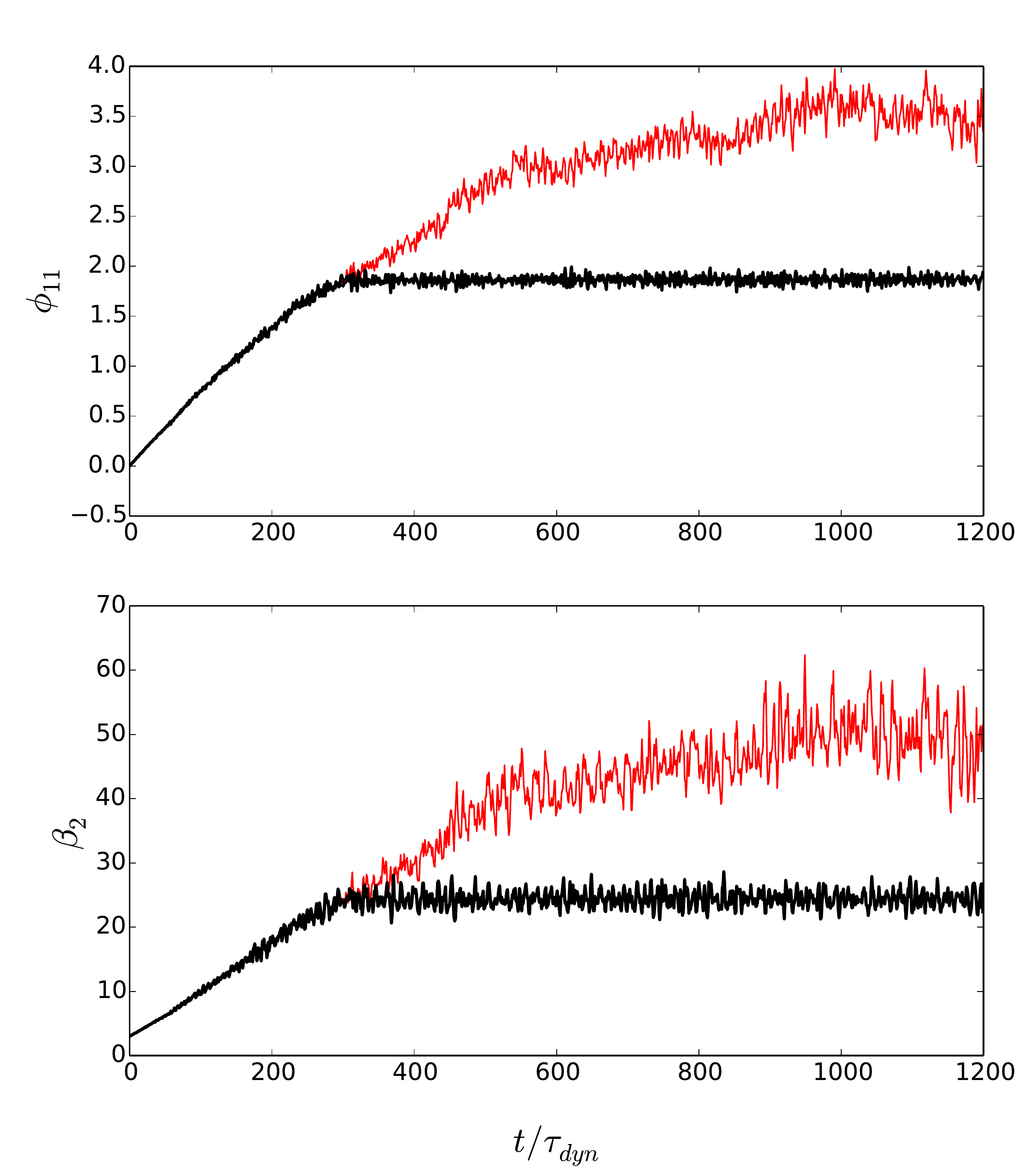}

\caption{Model A: Evolution of the entanglement parameter $\phi_{11}$ (top panel)  and of the dimensionless kurtosis $\beta_{2}$   averaged over $100$ realizations
 with $\gamma_{A}=0.1$ and  equilibrium initial conditions
for $N=1024$. The black curves correspond to a modified simulation in which the perturbation 
is switched off for  $t\geq 300\tau_{dyn}$. }
\label{fig:stop_diss}
\end{center}
\end{figure}

\subsubsection{Macroscopic evolution due to perturbation}

Fig.~\ref{fig:stop_diss} shows the time evolution of the entanglement parameter 
$\phi_{11}$ and the rescaled kurtosis $\beta_2$ (both defined above),  
averaged over $100$ realizations of equilibrium initial conditions
with $N=1024$ particles. 
As expected, at $t=0$, $\phi_{11}=0$ and $\beta_2=3$:  the equilibrium state 
phase space density is a separable function of space and velocity, 
and the velocity distribution has a Maxwell-Boltzmann shape. 
The red curve corresponds  in each case to the evolution with $\gamma_{A}=0.1$, 
while the black  curve is the evolution when this perturbation is
``switched off" at the time $t=300 \tau_{dyn}$, i.e., starting
from this time the evolution is that of the purely self-gravitating
system. Note that in line with what would be expected
from mean-field theory, the characteristic time for the
macroscopic evolution is about ten times shorter
than in the data in the previous figures, for
the case $\gamma_{A}=0.03$. We see that the 
evolution induced by the perturbation is 
through a continuum of QSS, i.e., at all times
the system remains very close to a stationary
{\it and stable} state of the Vlasov equation.
Given that the perturbation is weak ---
in the sense that it perturbs the system 
macroscopically on a time scale long 
compared to $\tau_{dyn}$ --- this is
indeed what one would expect.  However,
what is not evident, and to be underlined,
is that 
\begin{itemize}
\item the effect of the small perturbation is 
not, at sufficiently long times, perturbative: the 
system is clearly progressively driven very 
far from its initial state. This contrasts strongly
to the behaviour one would expect 
for familiar short range systems in which
a small perturbation would be expected 
(e.g. applying linear response theory) to
evolve to   an out of equilibrium state close to the
initial one, in the sense that all macroscopic
quantites are changed perturbatively. 
\item  the parameter $\phi_{11}$, which 
in absence of the perturbation would remain
stable at its initial value $\phi_{11} = 0$, 
evolves to a final value $\phi_{11} \approx 2-3$  
depending on $N$, i.e., towards
a state in which the correlation between
position and velocity is ever stronger.
The perturbation clearly drives the system
away from thermal equilibrium. 
 
\end{itemize}

\subsubsection{Validity of mean-field kinetic theory}

Let us now consider the degree to which the evolution in our 
simulations of the system are described well by the mean-field limit of the
kinetic equations derived above. The most basic prediction of
this theory is that, when we adopt the associated scalings of 
the parameters with $N$, we observe an evolution which is
independent of $N$.  The upper two panels of Fig.~\ref{fig:god_stoch_phy_all} 
show respectively  the evolution of $\phi_{11}$ and $\beta_2$,
in each case  averaged over $300$ realizations of the model 
with $\gamma_{A}=0.03$ starting from equilibrium initial  conditions,
and for the different particle numbers indicated: $N=128,256,512$. 
In both cases we observe that at sufficiently early times there
is a non-trivial evolution of the system which is very well
superimposed for the different $N$, indicating the validity
of the mean-field theory.  Further the evolution
of $\beta_2$ at these times agrees well with that
predicted by the mean-field kinetic theory at early
time,  shown as a straight line obtained using 
Eq. (\ref{kurtosis-evolutionKT}) with $f(x,v)$ 
taken equal to the initial thermal equilibrium
phase space density Eq. (\ref{eq:equivlasov}). 
At longer times however we see that the evolution
changes: for each $N$, the evolution breaks away, at a time 
scale which appears roughly to increase with $N$,  from the 
common behaviour, and shows on a similar time scale a 
tendency to reach a plateau, indicating in principle
the attainment of a stationary state. By measuring
the velocity and spatial distributions below we will
verify that this is indeed the case, and in so doing 
also find the explanation for the $N$-dependence
and the noisiness of the evolution of $\phi_{11}$ and 
$\beta_2$ at longer times: the stationary state to which
the system evolves is in fact one 
for which these particular macroscopic variables 
become ill defined in the mean field limit.
This is the case because these states are characterised by 
velocity and spatial distributions which have
slowly decaying power-law tails at long distances,
for which both $\langle |x| \rangle$ and
$\langle v^4\rangle$ diverge. Their values
in a finite simulation are then regulated by
the cut-off due to the finite particle number,
and are thus highly fluctuating. 
Shown in the lower panel of  
Fig.~\ref{fig:god_stoch_phy_all} is the
evolution of  $\langle |v| \rangle$,
which, in contrast, is a well defined 
quantity in the final state. In this case we
see that the evolution for different $N$ in
the mean-field scaling agrees well, and 
fluctuates little, right up to the time at which 
the stationary state is attained. The small
deviation at the latest times for $N=128$ 
is a result of the large fluctuations of
the total energy in this case 
(see Fig.~\ref{fig:god_stoch_tot_all} ).

\begin{figure}[ht]
\begin{center}
\includegraphics[scale=0.37]{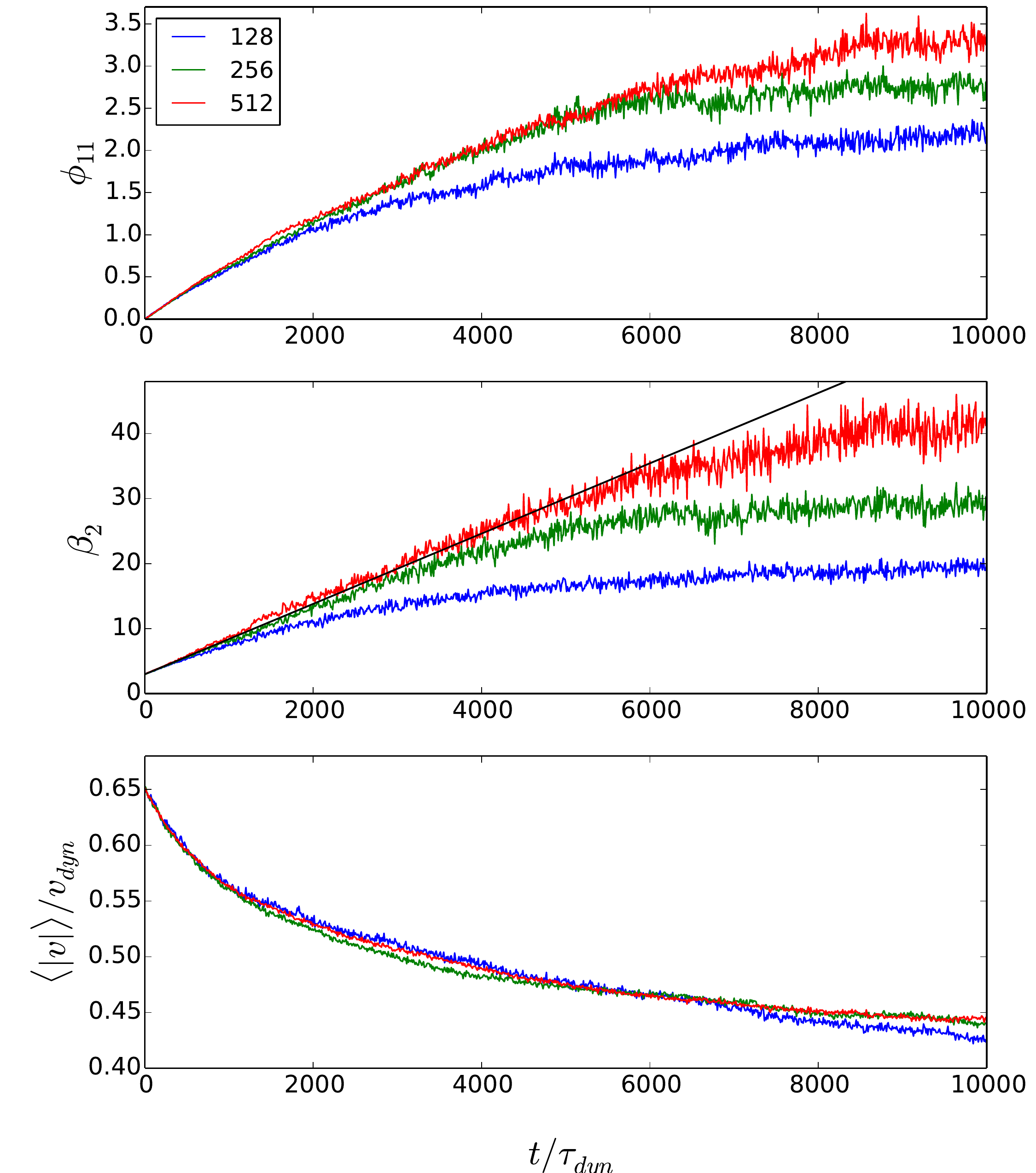}
\end{center}
\caption{Model A: Entanglement  parameter $\phi_{11}$,  dimensionless kurtosis of the velocity distribution $\beta_{2}$ and  $\langle |v|\rangle/v_{dyn}$, averaged over $300$ 
realizations, as functions of time $t/\tau_{dyn}$   for $\gamma_{A}=0.03$ and   $N=128,256,512$ (from top to bottom).}
\label{fig:god_stoch_phy_all}
\end{figure}

\subsubsection{Dependence on initial conditions}

 Fig.~\ref{fig:phi11_rea} shows  the evolution of $\phi_{11}$ and
 $\langle |v| \rangle/v_{dyn}$  for $\gamma_A=0.03$ and $N=512$ and with three 
different initial conditions: thermal equilibrium, and
two rectangular waterbags, $R_0=1$ and $R_0=0.01$. 
The number of realizations are $300,100, 200$ respectively. At long times,
these quantities  evolve towards  the same mean  value, independently of the initial conditions.
Note that for  $\phi_{11}$   
fluctuations increase with time and are associated with the long tails of position and velocity distributions.

In the  presence of the perturbation, the system goes
 at  long times to a non equilibrium stationary state 
 which is an attractor of the dynamics, i.e., the perturbed
dynamics of this long-range system has a  ``universal''
stationary state. 
 
 \begin{figure}[h!]
 \begin{center}
\includegraphics[scale=0.37]{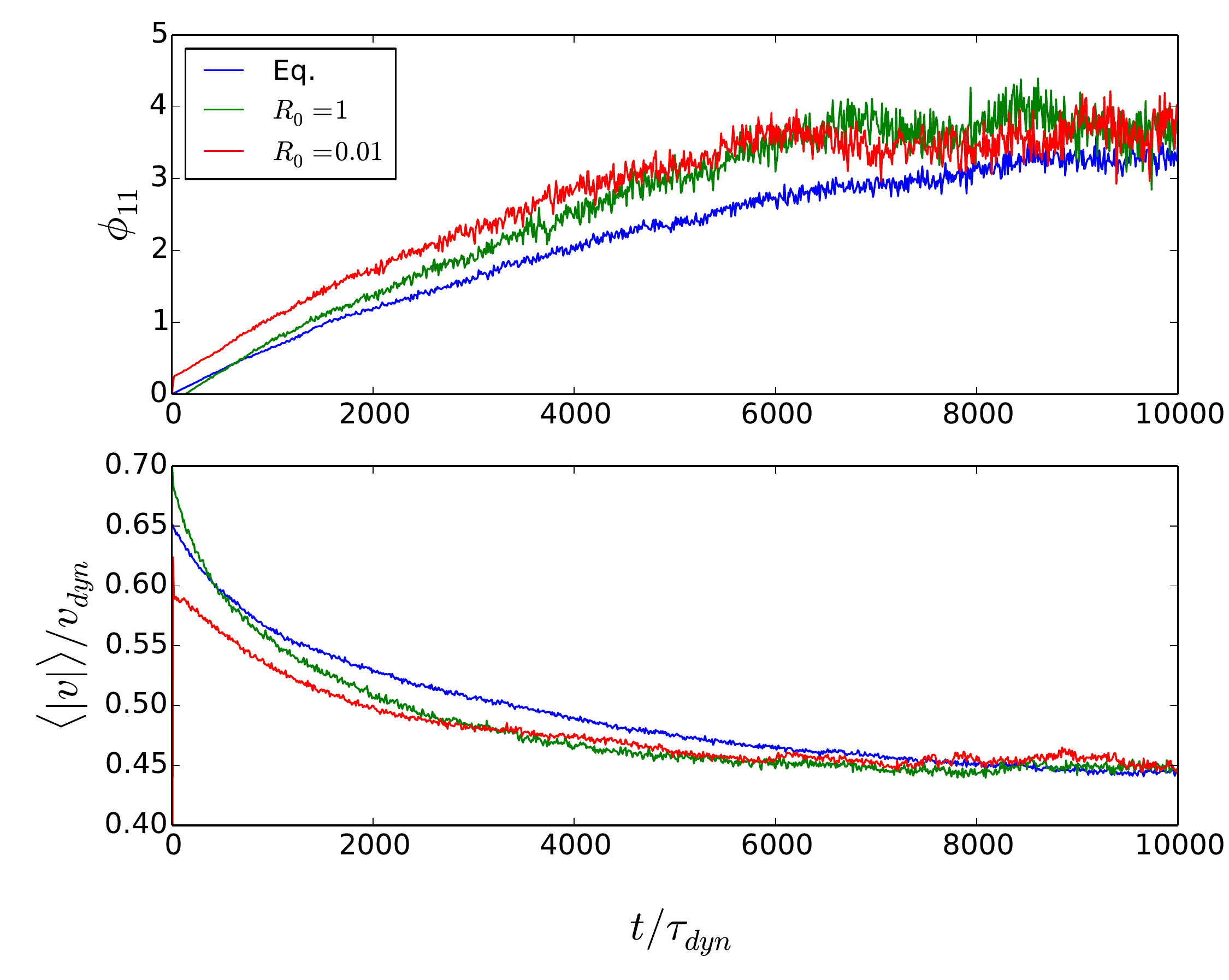}
 \end{center}
 \caption{Model A: $\phi_{11}$ (top) and of $\langle |v| \rangle/v_{dyn} $ (bottom) as a function of time $t/\tau_{dyn}$, 
 with $\gamma_A=0.03$ and $N=512$ for different initial conditions:
 thermal equilibrium (averaged over 300 realizations), rectangular waterbags 
 with $R_0=1$ (100 realizations), and $R_0=0.01$ (200 realizations).}
 \label{fig:phi11_rea}
 \end{figure}

\subsubsection{``Universal'' final state}
Let us consider now the properties of this apparently stationary state,
and check in particular whether the phase space distribution is indeed
stationary and the same in the different cases.

Fig.~\ref{fig:stoch_pdf} shows the velocity probability distribution   averaged 
over $100$ realizations, for  $\gamma_{A}=0.1$ and $N=1024$ and with an initial  thermal 
distribution (black curve in the left panel). The analogous spatial distributions are shown in 
Fig.~(\ref{fig:stoch_pdf_x}) for the same data. In both cases the relevant variable has been 
normalised to the square root of its variance at the given time.  
Except for the initial distribution (cf. Eq. \ref{eq:equivlasov}) and the next time plotted ($t=500\tau_{dyn}$), all
the curves are thereafter very well superimposed. The tail of the corresponding velocity distribution is
well fitted by a simple power-law behaviour $\sim v^{-\kappa}$, with $\kappa \approx 3$, while that
of the spatial distribution by $\sim x^{-\epsilon}$, with $\epsilon \approx  2$. We find the same
behaviours for our different initial conditions, and for the different $N$ in the range we 
have considered. Further we note that we find these behaviours to be stable even if we 
extend our analysis to data at longer times, in which the fluctuations of energy become large.
Thus the state appears to a very robust attractor even when the energy can vary considerably.
As anticipated above, these asymptotic behaviours of the evolved system explain why we 
observed the strongly $N$ dependent behaviours of the parameters $\phi_{11}$ and $\beta_2$: 
for such asymptotic behaviours of the velocity and space distribution these quantities are divergent, 
and thus in practice, when measured in a system with a finite number of particles, they are 
dominated by the contribution from just a few of the highest energy particles. 

We note that the velocity distribution observed is very similar to that found for 
the original purely granular model (i.e.without gravity) \cite{Barrat2001}. As in
this case one must in fact suppose that $\kappa >3$ to ensure that the 
kinetic energy (proportional to the velocity dispersion) of the state be 
finite.

\begin{figure}[ht]
\begin{center}
\includegraphics[scale=0.37]{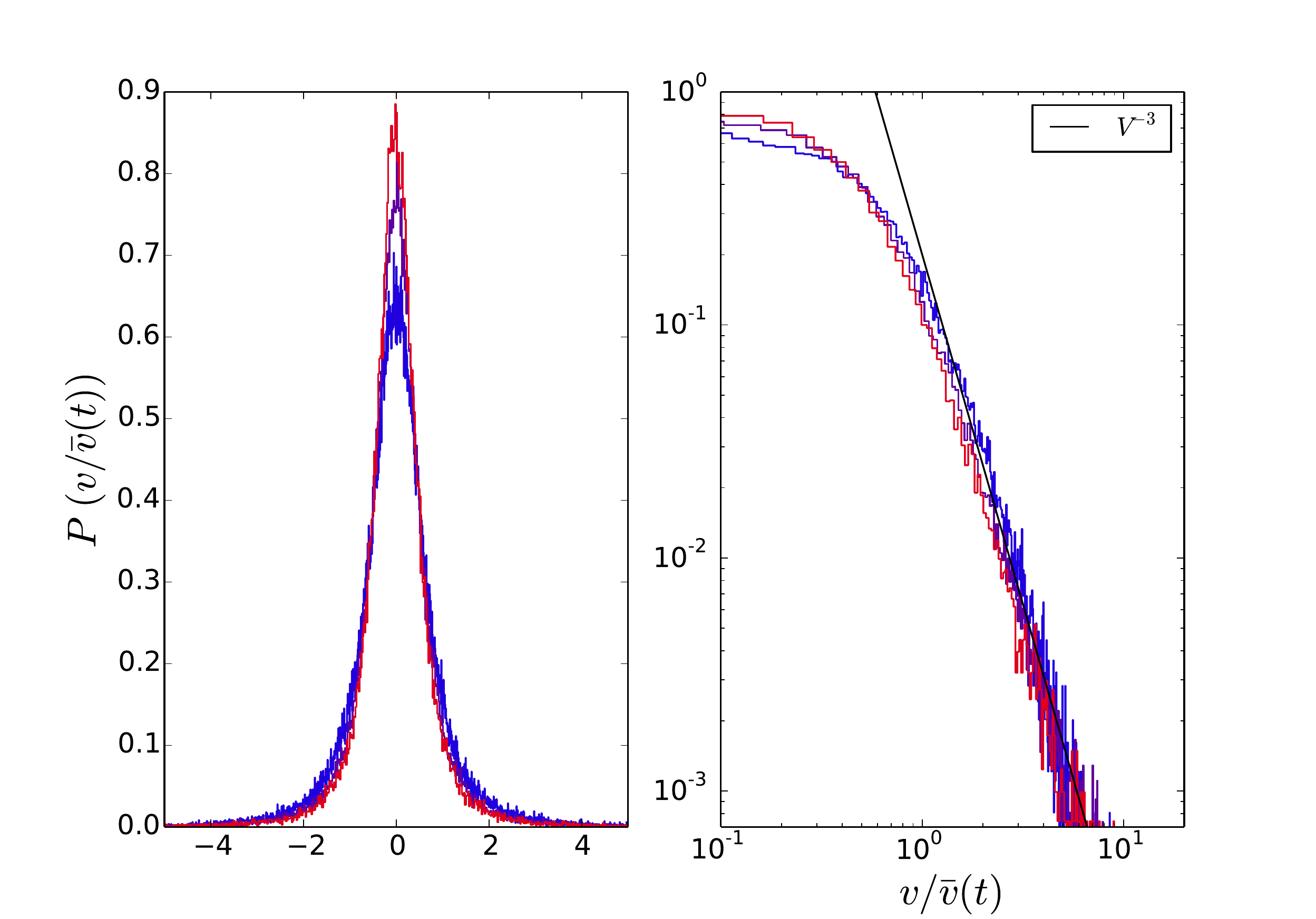}
\caption{Model A: Velocity distribution  at different times ($t=0,500,1800,2500,3700\tau_{dyn}$) 
averaged over $100$ realizations   for $\gamma_{A}=0.1$, 
 $N=1024$ with initial thermal distribution.  Velocities are normalised by $\overline{v}(t)$, 
the standard deviation  of the velocity distribution. The left panel is a 
linear plot and the right panel a log-log plot.}
\label{fig:stoch_pdf}
\end{center}
\end{figure}

\begin{figure}[ht]
\begin{center}
\includegraphics[scale=0.37]{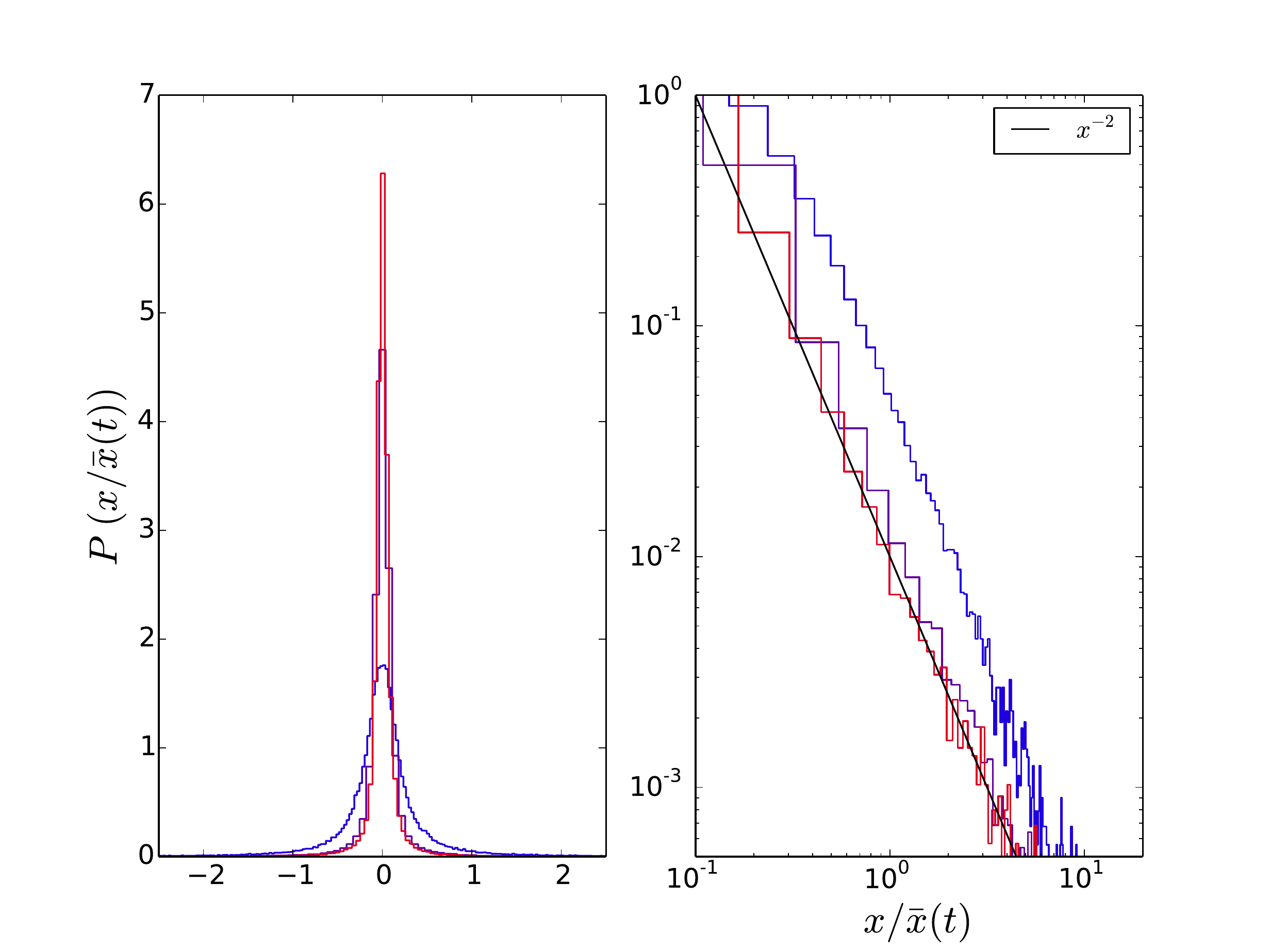}
\caption{Model A: Position distribution at different times ($t=500,1800,2500,3700\tau_{dyn}$) averaged 
over $100$ realizations  for $\gamma_{A}=0.1$, $N=1024$ with  initial thermal distribution.
 Positions are normalised by $\overline{x}(t)$,  the standard deviation  of the position distribution. The left panel is a linear plot and the right panel a log-log plot.}
\label{fig:stoch_pdf_x}
\end{center}
\end{figure}

\subsection{Model B}

We have seen that the contribution from collisions in Model B 
is characterized in the mean field limit by the dimensionless 
parameter $\gamma_B$, and the velocity scale $v_0$, with
both being held fixed in the mean-field limit. As the mean-field
scaling leaves invariant also the characteristic velocity
$v_{dyn}$ defined in Eq.~(\ref{def-dynvel}), we can
define the dimensionless ratio
\begin{equation}
u_B=\frac{v_0}{v_{dyn}}
\end{equation} 
which also remains fixed in the mean field limit.
We then can characterize our simulations by the dimensionless
parameters $\gamma_B$, $u_B$ and $N$, and the results will 
then be $N$-independent at sufficiently large $N$ if
the mean-field treatment is valid.

\subsubsection{Macroscopic evolution due to perturbation}
\label{Macroscopic evolution due to perturbation}

\begin{figure}[h!]
\begin{center}
\includegraphics[scale=0.37]{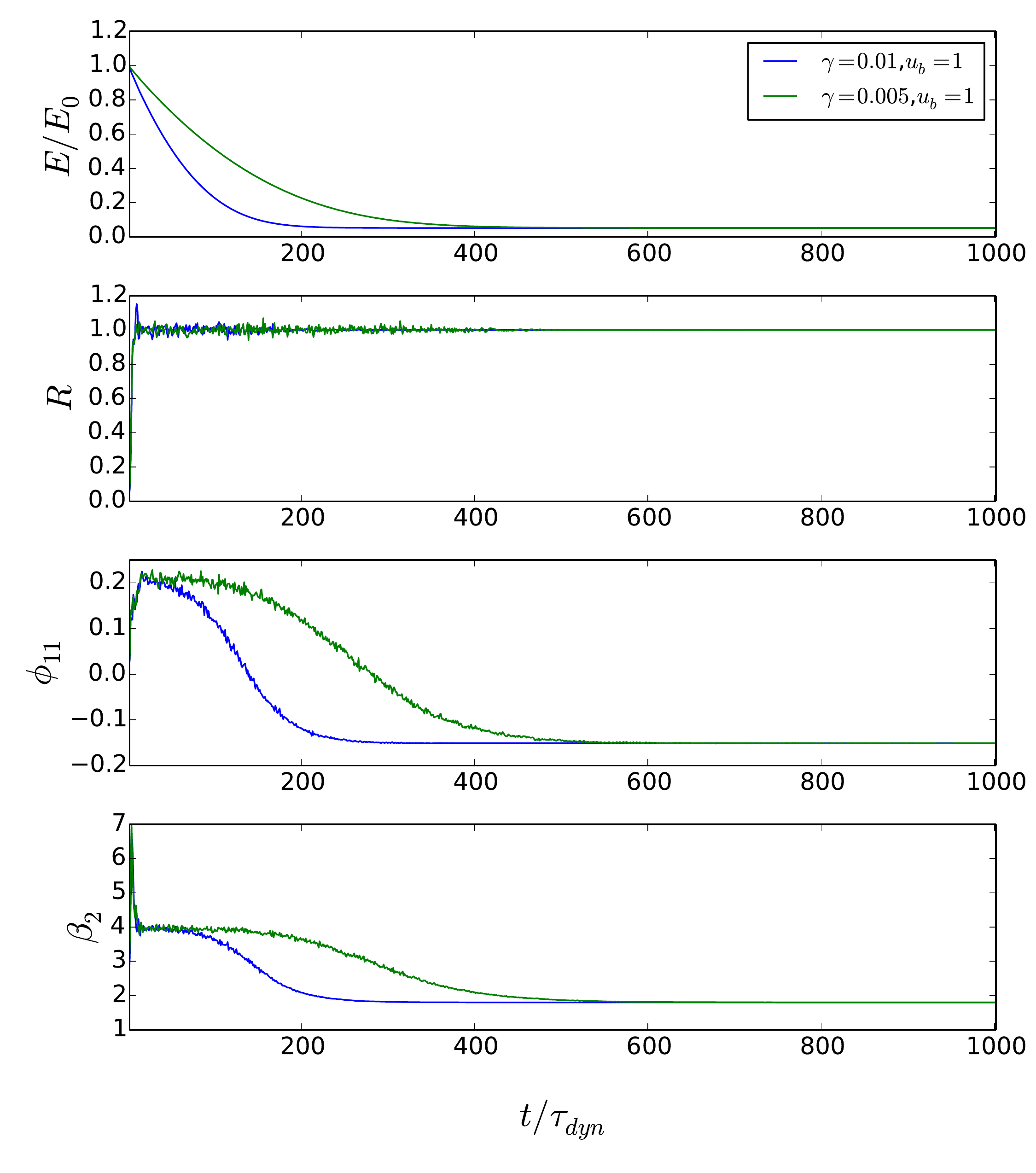}
\caption{Model B: Dimensionless energy $E/E_0$ ($E_0$ is the initial energy), virial ratio $R$, entanglement parameter $\phi_{11}$ and dimensionless kurtosis $\beta_2$
versus $t/\tau_{dyn}$ for $u_B=1$, $\gamma_B=0.01,0.005$, $N=512$ and initial rectangular waterbag conditions $R_0=0.01$. Simulation results are averaged  over $20$ 
realizations.}
\label{fig:BRS_GOD_R0_R1}
\end{center}
\end{figure}

Fig.~\ref{fig:BRS_GOD_R0_R1} shows results for the evolution of the 
dimensionless energy $E/E_0$  ($E_0$ is the initial value), of the virial ratio $R$,  $\phi_{11}$ and $\beta_2$,  
for $N=512$, $u_B=1$ and   rectangular waterbag 
initial conditions  with $R_0=0.01$.  the two curves correspond to the  different values of $\gamma_B=0.01,0.005$ with an average over $20$ realizations. 
We observe that, as expected, the 
system reaches virial equilibrium on a time scale $\sim 10 \tau_{dyn}$
and remains, to an extremely good approximation,
virialized thereafter. Compared to model A, the
finite $N$ fluctuations are extremely small. 
As we will see in further detail below, this is 
a result of the presence of a well defined energy 
scale in the model to which the system is efficiently 
driven.  
Thus the system evolves on a time scale
$\sim \tau_{dyn}/ \gamma_B$ as expected
from the mean-field kinetic theory, through
a continuum of QSS. Indeed, to test this 
conclusion, we have performed again
simulations in which we ``turn off" the perturbation
at different times. We find, as in model A, that the
macroscopic parameters remain essentially
frozen at their values at this time.

For the chosen value of $u_B=1$ the simulations start 
with an energy which turns out to be about an order 
of magnitude larger than the energy in the stationary
state. This means that the characteristic velocities
are initially so large that most collisions are 
inelastic and the evolution depends little on
the presence of the term depending on $v_0$. 
In this case the evolution is then well
approximated by the case of purely inelastic 
collisions which we have studied in \cite{Joyce2014}.
For smaller values of $\gamma_B$ than those
shown here the validity of this approximation
is sufficiently extended in time so that one can
see the presence of an approximate plateau
in $\phi_{11}$ corresponding to the ``scaling
QSS''  derived in this work.

\subsubsection{Dependence on initial conditions}
\begin{figure}[t!]
\begin{center}
\includegraphics[scale=0.40]{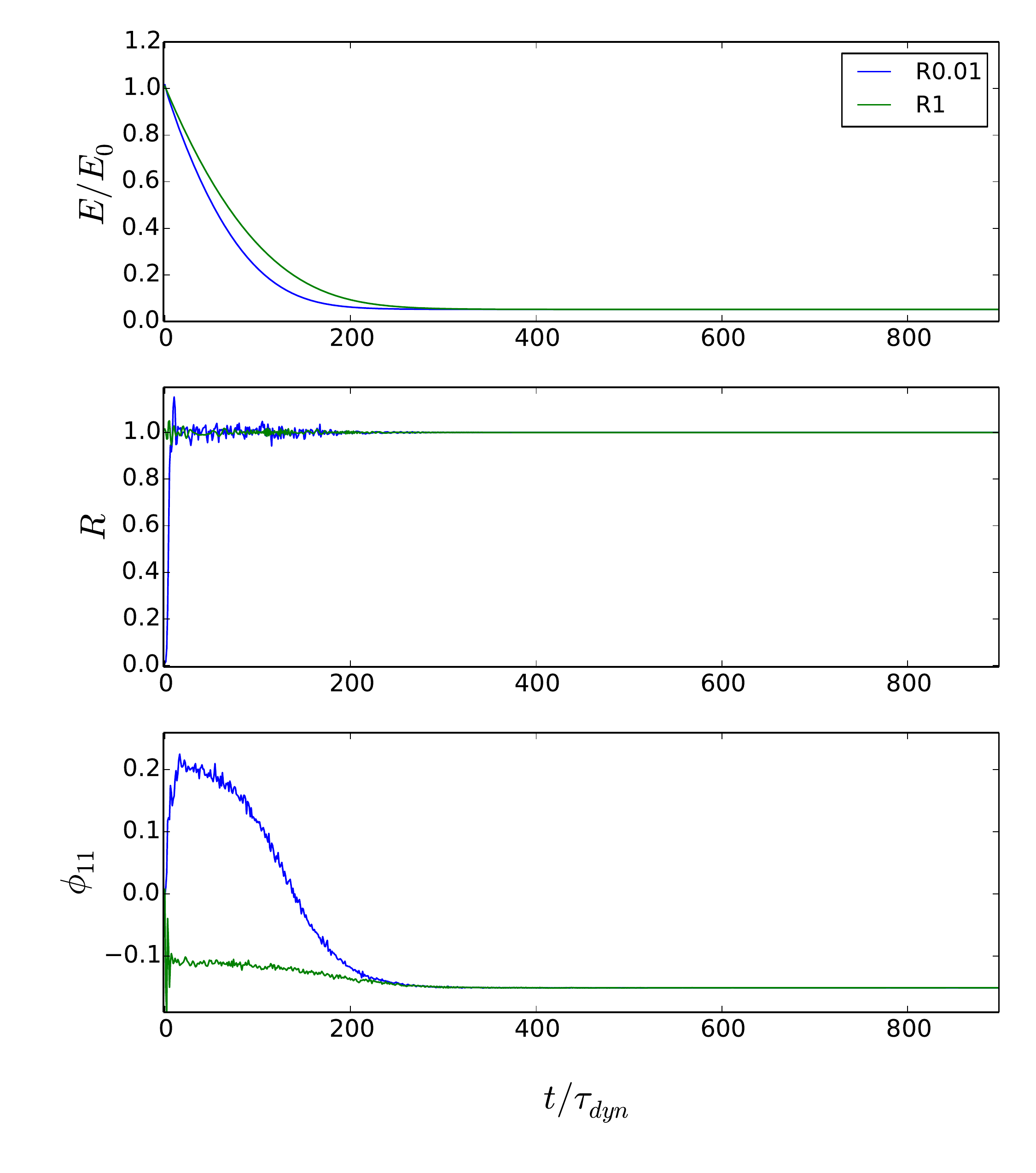}
\caption{Model B:  $E/E_0$ (top) and $\phi_{11}$ (bottom)
 versus time $t/\tau_{dyn}$ for $N=512$, 
$\gamma_B=0.01$, $u_B=1$ with rectangular waterbag initial conditions, $R_0=0.01, 1$.} 
\label{fig:BRS_diff_IC}
\end{center}
\end{figure}

Fig.~\ref{fig:BRS_diff_IC} shows the evolution of the 
energy $E/E_0$, $R$ ,
and $\phi_{11}$ for 
$\gamma_B=0.01$, $u_B=1$, $N=512$  and for  two different  rectangular waterbag initial conditions, $R=0.01$ and $R=1$).
We see, as indicated by the
behaviour of $\phi_{11}$, that each of the two initial 
conditions initially evolves to a quite different QSS, but 
then on the longer time scale both converge towards 
an identical value of $\phi_{11}$. That this indeed 
corresponds to evolution to the same final state is
confirmed, as we will detail further below, by study
of the final configuration in phase space.

\subsubsection{Mean-field limit of kinetic theory}

To test the validity of the mean-field limit derived in Section \ref{sec:modelI}, 
we have run sets of simulations for the same initial conditions with fixed values
of $\gamma_B$ and $u_B$,  but different values of the particle number $N$.  
Fig.~\ref{fig:BRS_GOD_energy} shows the evolution 
of the energy $E/E_0$ (top ) and  $\phi_{11}$ (bottom)
for 
$\gamma_B=0.01$, $u_B=1$, and  for rectangular waterbag initial conditions, $R_0=0.01$ with different system sizes, $N=128, 256,512$.
We see in the evolution of the energy an almost perfect superposition
of the curves, indicating thus an $N$ independent evolution corresponding
to the mean-field limit. 

\begin{figure}[h!]
\begin{center}
\includegraphics[scale=0.37]{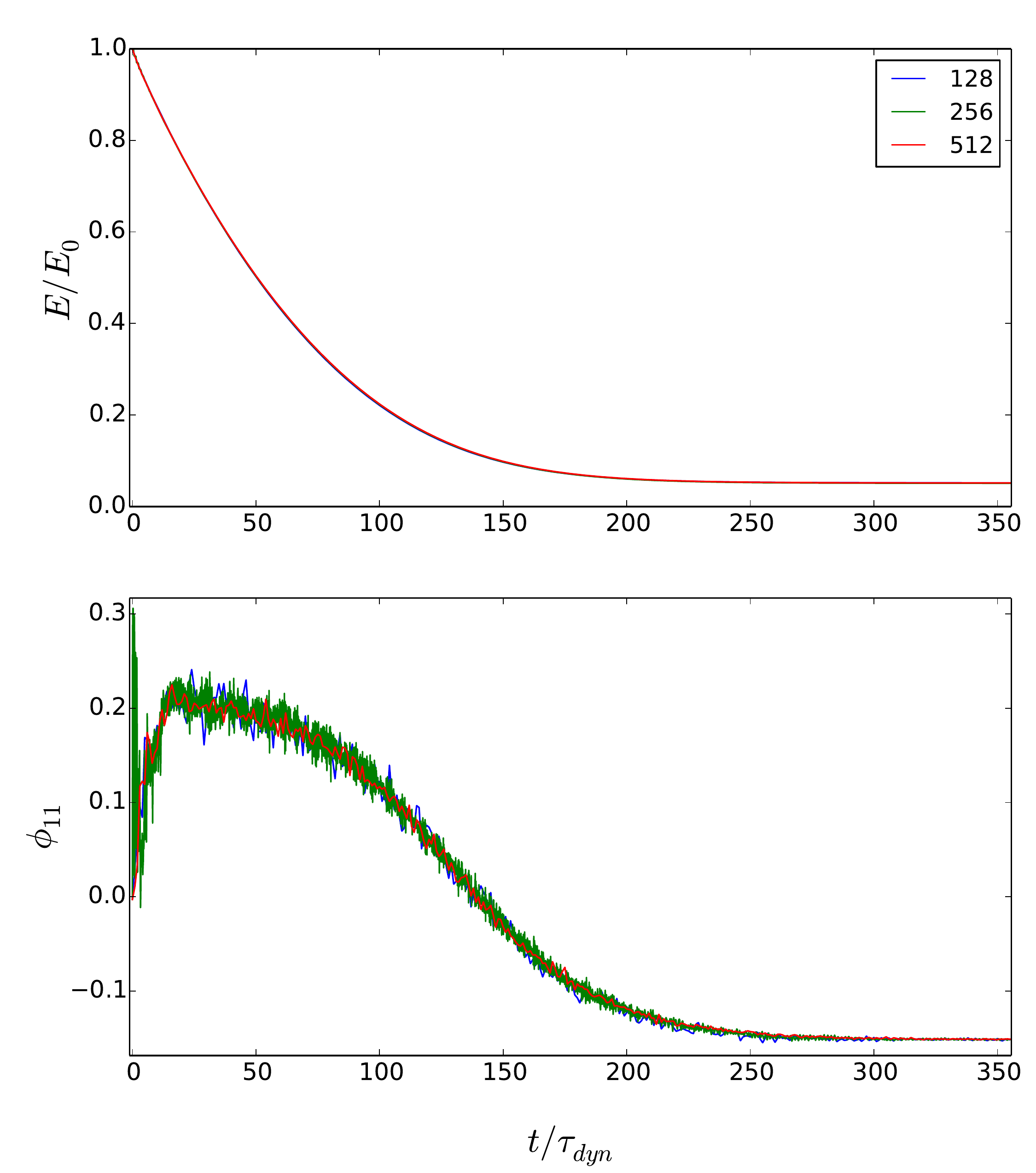}
\caption{Model B: $E/E_0$ (top) and $\phi_{11}$ (bottom) for 
$\gamma_B=0.01$, $u_B=1$, and  for rectangular waterbag initial conditions, $R_0=0.01$ with different system sizes, $N=128, 256,512$. }
\label{fig:BRS_GOD_energy}
\end{center}
\end{figure}

\subsubsection{Properties of ``universal" final state}

We finally consider in greater detail the properties of the apparently 
very well defined final state to which the system is driven very efficiently 
in Model B. 
Fig.~\ref{fig:BRS_PhaseSpaceR01} shows snapshots, at the 
indicated times, of the phase space of particle positions in 
dimensionless units (where $x_0=v_0\tau_{dyn}$) in $20$ realizations 
with $N=128$ of $R=0.01$ waterbag initial conditions, for a model
with $\gamma_B=0.01$ and $u_B=1$.  The phases of the evolution, 
already evident in the evolution of the energy and $\phi_{11}$ 
as discussed above, are again clearly visible. However the phase space 
plot reveals that, from the time (here about $150 \tau_{dyn}$) at which 
the macroscopic diagnostics indicate the establishment of the 
stationary states, and do not themselves appear to evolve
anymore, there is a further non-trivial evolution in phase 
space of the {\it microscopic} particle distribution: the particles 
progressively ``aggregate" onto distinct separated 
curves. A study of the particle energies shows that 
they are, to a very good approximation, constant on
each curve, and take well separated values on each curve
i.e. they are effectively discretized.

\begin{figure}[h!]
\begin{center}
\includegraphics[scale=0.33]{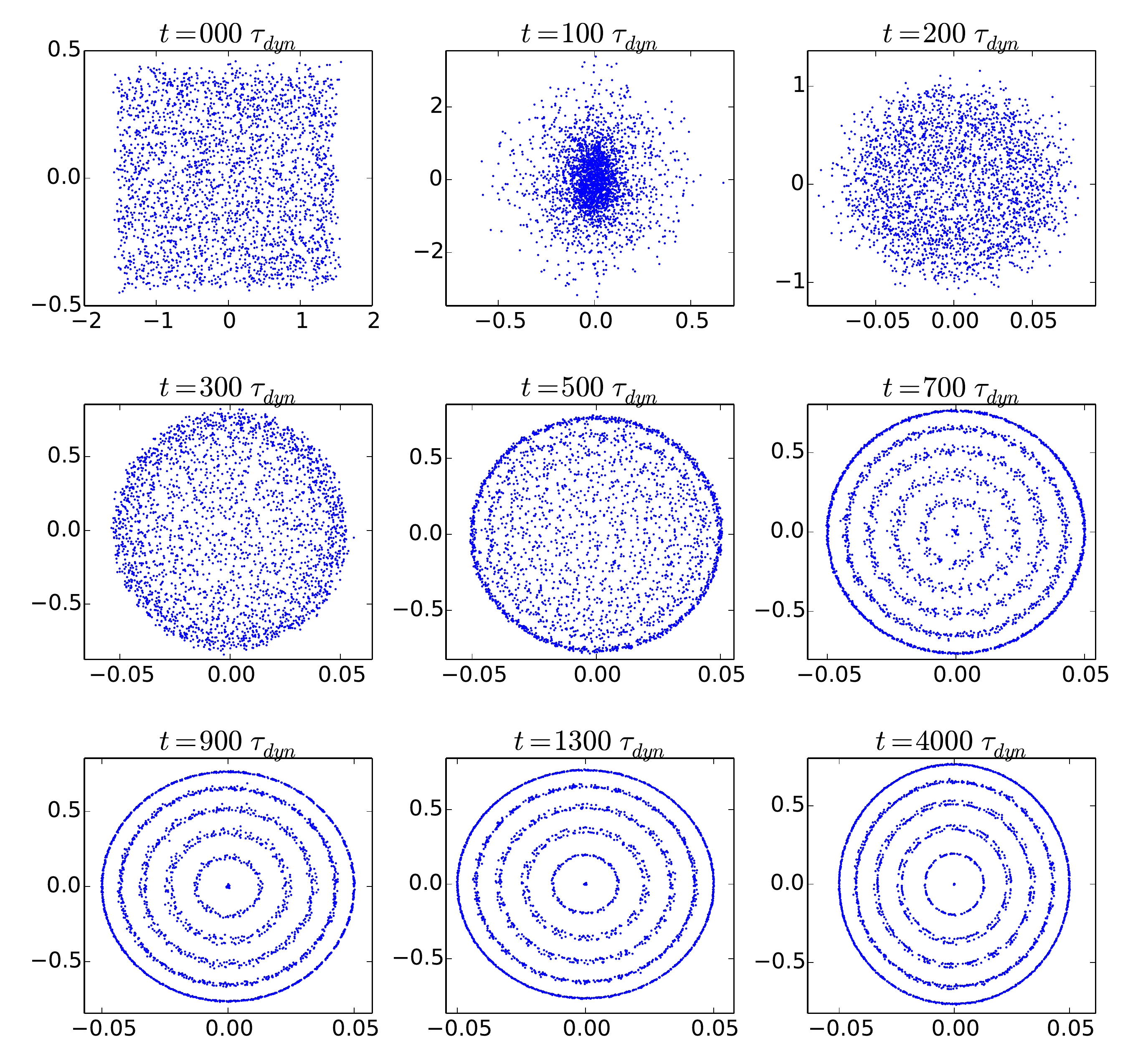}
\caption{Model B: Phase space snapshots in dimensionless units ($x/x_0$ and $v/v_0$)  
 for $N=128$,  $\gamma_B=0.01$,  $u_B=1$,  rectangular waterbag
initial conditions,  $R_0=0.01$ and  $20$ realizations   at different times $t$
ranging from $t=0$ to $t=4000 \, \tau_{dyn}$).}
\label{fig:BRS_PhaseSpaceR01}
\end{center}
\end{figure}
\begin{figure}[h!]
\begin{center}
\includegraphics[scale=0.42]{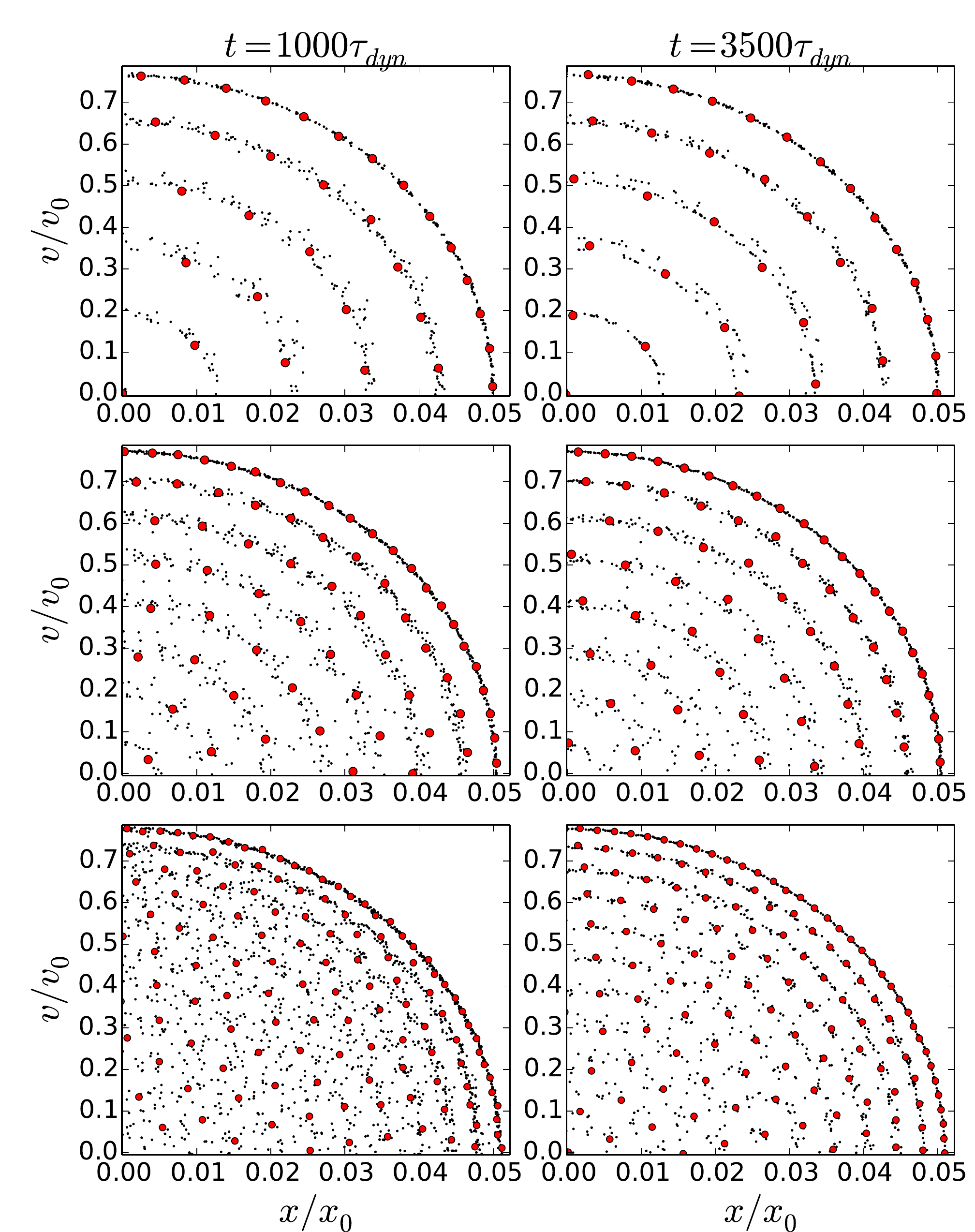}
\caption{Model B: Phase space  snapshots in dimensionless units ($x/x_0$ and $v/v_0$)  
for  
$\gamma_B=0.01$, $u_B=1$,   rectangular waterbag
initial conditions, $R_0=0.01$, $20$ realizations and at   two different times
$t=1000 \tau_{dyn}$ (left panels), $t=3500 \tau_{dyn}$ (right panels).  We compare three different system sizes $N=128,256,512$ (from top to bottom). 
Red dots correspond to one chosen realization, small black dots  correspond to the other realizations.}
\label{fig:BRS_PhaseSpace_rea}
\end{center}
\end{figure}

Fig.~\ref{fig:BRS_PhaseSpace_rea} shows phase space
configurations at two times ($t=1000 \tau_{dyn}$ on left,
$t=3500 \tau_{dyn}$ on right) for $20$ realizations of the same 
waterbag initial condition, and the same $\gamma_B$ and 
$u_B$, as in the previous figure, for $N=128$ (upper panels), 
$N=256$ (middle panels) and $N=512$ (bottom panels). In each plot 
the particle positions for a single chosen realization are also plotted 
as red stars.  These plots show clearly that in all cases 
the system evolves towards a highly ordered distribution, in 
which the particles are not only on ``shells" in phase space 
as noted above, but also have highly ordered 
positions along these shells i.e. the relative phases of the 
particle' motions on the shells are fixed in time and 
completely coherent. Further the time scale to attain 
the final state appears to grow strongly with $N$ 
(in units of $\tau_{dyn}$):  the $N=128$ simulations have
already attained the completely ordered state at 
$t=1000 \tau_{dyn}$, the $N=256$ simulations are close
to attaining it for $t=3500 \tau_{dyn}$, while the 
$N=512$ systems are still evolving towards it
at this later time. 

This evolution of the system towards the ``ordered"
state thus occurs on a time scale which diverges
when we take the mean-field limit. Indeed it is
an evolution intrinsically characteristic of the 
finite $N$ system. However, as far as we can
determine, the macroscopic properties of the
system are unchanged by the corresponding
microscopic evolution e.g. the total energy 
and parameters $\phi_{11}$ and $\beta_2$
do not evolve on average. Thus the QSS,
which corresponds to the phase space density 
in the infinite $N$ limit, appears to remain
unaltered. This behaviour can be contrasted with the 
relaxation to thermal equilibrium of QSS attained
in the purely self-gravitating model. This relaxation
is also driven by finite $N$ effects, but it causes 
the system to evolve (on a time-scale $\sim N \tau_{dyn}$)
through a family of QSS until it finally equilibrates.

As mentioned in Section \ref{Macroscopic evolution due to perturbation} 
above, we have verified that the intermediate states the system
evolves through from the time it virializes are indeed a family of QSS
of the purely self-gravitating model: when we turn off the
perturbation any time after virialization, the system's macroscopic
properties do not change on mean field time scales. For  
a few cases with $N=128$ we have evolved them long 
enough to see that, as expected, they then evolve
towards thermal equilibrium on a time scale 
$\sim N \tau_{dyn}$. Performing the same experiment
starting from a time at which the same system has
had time to evolve to the ``ordered" microscopic state
we find an intriguing result: this ordered microscopic
state remains unchanged under the purely gravitational
evolution, not only on the mean-field time scale 
($\tau_{dyn}$ and $\tau_{dyn}/\gamma_B$) but 
even on the time scale much greater than $N \tau_{dyn}$.
Thus the microscopic state attained at long times
in presence of the perturbation appears to be a 
periodic or quasi-periodic solution of the
pure gravitational $N$-body system, and 
to belong to a stable island in the $N$-body
phase space which leads to a breaking of
ergodicity. We will investigate further both
the dynamics giving rise to and the properties
of these intriguing ``ordered" states of the 
$N$-body system in future work.
  
\section{Discussion and conclusion}

We have investigated the effects on the dynamics of long-range interacting systems 
of a class of ``local internal'' perturbations through the study of a canonical one dimensional 
toy model subjected to such perturbations.  More specifically we consider two perturbations 
inspired by granular studies of the dynamics of a one dimensional self-gravitating
system considering momentum conserving, and energy violating, collisions which are 
designed so that they can, nevertheless, conserve energy in an average sense.
Our main focus has been on the question of how the characteristic non-equilibrium 
stationary states or QSS of the long-range system are affected by these
perturbations. We consider the case that these perturbations are 
{\it weak}, in the sense that the time scale on which they affect the
system macroscopically are long compared to the time scales
characteristic of the dynamics of the mean gravitational field.

We have derived first kinetic equations for both models which describe the 
system's evolution in a large $N$ mean-field and quasi-elastic limit.  
Our numerical study of the models shows that this limit describes
well the macroscopic evolution due to the perturbations at sufficiently
large $N$,  but at given $N$ we see always also at sufficiently
long times  evolution in both models which are finite $N$ effects
not captured by the mean field treatment.  In model A such effects 
are manifest in large excursions of the energy at longer times,
and in model B in the appearance of a highly ordered 
microscopic phase space distribution.
Within the regime of validity of the mean field approximation
both models show at longer times evolution towards an apparently unique virialized state.
This state is not the thermal equilibrium of the isolated model, and
indeed is typically ``further  away'' (in terms of correlations measured by  $\phi_{11}$) from the thermal equilibrium.
Therefore we observed compelling evidence for the establishment of an attractor ``universal'' non-equilibrium 
stationary state in both models.
Despite the stochasticity of the dynamics, explicit in model A,  the system shows no tendency to relax toward equilibrium
(as observed for example in the HMF model \cite{Gupta2010PRL})

Both perturbations, which act microscopically and locally, thus completely modify the global 
organisation of the system: they drive the long-range system far 
from the QSS it is in initially (due to mean-field relaxation from the
initial conditions on time scales significantly shorter on which
the perturbations act). In this sense the QSS are not robust to such perturbations, and are modified
macroscopically as soon as the perturbation starts to act. However, the evolution which results 
is through a succession of virialised states which are stationary solutions of the Vlasov equation. In both
models the system is then driven finally to a non-equilibrium stationary state (NESS),
which is itself also a QSS of the unperturbed long-range system.
This final state does not depend on the initial conditions, but does depend strongly on 
the details of the perturbation. Indeed in the two models we have considered the final state 
has completely different properties, with notably in model A a power-law decaying 
space and velocity distribution compared to a phase space distribution
with compact support for Model B. Thus the perturbation, albeit apparently
very weak, turns out to completely dominate and determine the behaviour
of the long-range system. This contrasts dramatically with the effect 
such a weak perturbation would have on a short range system which 
relaxes efficiently to thermal equilibrium: in this case the perturbation
would indeed just  perturb slightly this equilibrium.

It is interesting to compare our results with related previous work in the literature.
In \cite{Gupta2010PRL,Gupta2010b} the stochastic perturbation applied to the long-range system 
(the HMF model) permutes the momenta of triplets of particles chosen
randomly in the system, and drives the system to relax to thermal equilibrium
efficiently. Applied to a one dimensional self-gravitating system, we have
checked that we observe the same behaviour. Indeed it suffices in this case 
to consider exchanges of the velocities of randomly chosen pairs of particles
because the QSS are inhomogeneous. This drives the system efficiently to
equilibrium because it destroys directly, because of the non-locality of
the perturbation, the entanglement in space and velocity of the phase 
space distribution.   In the models we have presented here the perturbation, as 
we have underlined, has the property of being \textit{local}. This means that the 
instantaneous change of the  velocity distribution induced by the collisions 
depends on the local properties, and 
as these typically vary in space in a non-trivial manner there is no reason
to expect the system to evolve towards the same  velocity distribution everywhere. 
On the contrary, as we have seen, the system tends to
evolve towards configurations in which the space and velocity distributions are 
ever more strongly entangled, until a stationary state is reached in which the spatial
organisation induced by the long-range forces ``compensates" the local
modification of the velocity distributions by the perturbations.

Our study is complementary also to that of references ~\cite{Nardini2012,Nardini2012a} 
in which the effect of an \textit{external} stochastic force acting on a long-range system
is studied. Indeed we can consider the perturbations we have introduced at particle
collisions as stochastic forces, with the difference that they are \textit{internal},
i.e. they are determined by the instantaneous microscopic state of the system 
itself. As we have mentioned, our models are thus appropriate to model the effects, 
for example, of additional short-range interactions at play in the system, while that
of  \cite{Casetti2012} models the effects of interactions with matter external to 
the system. For their treatment with kinetic theory, our models admit a considerable
simplification compared to that required for the models of ~\cite{Nardini2012,Nardini2012a}: as 
we have seen, we obtain in both our models a non-trivial kinetic theory which
includes the effect of the perturbation in the mean-field limit, i.e., by neglecting
two point correlations in phase space. As described in ~\cite{Nardini2012,Nardini2012a}
a non-trivial large $N$ limit for the evolution induced by the external perturbation
is obtained going beyond the mean field limit, and specifically can be obtained
by including non-trivial two point correlations. The reason for this difference is
that the action of the external forces on the system depend crucially on the
spatial correlations of these forces, and their effect on the evolution cannot
be described self-consistently without incorporating the resultant correlations
in the perturbations to the phase space distributions. While \cite{Nardini2012,Nardini2012a}
can obtain a range of different behaviours from the external stochastic 
forces --- ranging from thermalisation of the system to out of equilibrium
states characterised by intermittency --- our models display the simpler 
phenomenology of attractive non-equilibrium steady states we have 
described. 

We have constructed our models so that they either conserve energy on average (model A, in the large $N$ limit) or
can attain states in which energy is stationary (model B). When considering perturbations to such systems, there
is no reason in general to expect them to have such a property. What would we expect the effect notably of net energy
dissipation or injection to be? In \cite{Joyce2014} we have considered a simple classes of perturbations which 
dissipate energy, and found that they admit what we have called ``scaling QSS" . These are states of such systems 
in which the dissipation of the energy leads simply to an evolution in which the phase space density remains unchanged
other than to an overall rescaling of its characteristic size and velocity.  This study suggests that in models
like those considered here, but including a constant energy dissipation, one might expect to see established
an ``attractive scaling QSS" i.e. evolution to a unique phase space distribution in rescaled variables
reflecting the dissipation of energy. Indeed we note that in model A  extrapolated to the very long time
scales where the macroscopic energy strongly fluctuates due to finite $N$, we have found that,
in suitably rescaled coordinates, the system's velocity and space distributions remains very stable,
with notably the same power law tails measured in the mean field regime.    

We have studied here only two very specific and simple models and further study will be required
to determine how generic to long-range interacting systems the interesting behaviours we have
observed are --- in particular the evolution towards a unique stationary state which is a QSS of the
unperturbed system ``selected" by the perturbation. Nevertheless, on the basis of what we 
have observed, we believe it is reasonable to anticipate that such behaviour may indeed be 
common to many such systems. Isolated long-range systems admit an infinite 
number of QSS, and which of these states the system relaxes to on mean field time scales
is determined by the initial conditions, and depends on them in general in a way which 
is extremely complex (see e.g. \cite{Lynden-Bell1967,Lynden-Bell1999,PhysRevE.75.011112,Joyce2011a,Pakter2011,Benetti2014}). The application of a weak perturbation
to the system can be understood as providing a breaking of the degeneracy of the infinite number
of QSS which drives the system to a QSS which is invariant under its own action i.e. in which, 
in our models,  the collision term induced by the perturbations is zero. As the perturbation will
generically violate all the conservation laws (Casimirs) of the Vlasov dynamics, there would
appear to be no reason why the perturbed dynamics cannot explore the full space of QSS 
accessible from a given starting energy and mass, and thus ``find" the stable state 
starting from any initial condition.  Further generically we would not expect this final state to 
be the thermal equilibrium of the system (which is a particular QSS): as we have underlined,
unless the perturbation applied locally to the velocity distribution tends to drive the
system everywhere to the same velocity distribution, we expect the long-range force
to give rise to a stationary state in which the velocity and space distributions are
entangled in a manner characteristic of non-equilibrium QSS.

We thank Fran\c{c}ois Sicard for the original code for the 1D self-gravitating system,  
Thomas Epalle for results obtained on thermalisation of this system with a random two particle exchange 
algorithm, and Andrea Gabrielli for very useful discussions about Model B.


%
\end{document}